# Scaling in Kinetics of Supercooled Liquids

**B. Zhang[a*], D. M. Zhang[b*], D. Y. Sun[a§] and X. G. Gong[b§]**

[a]School of Physics and Electronic Science, East China Normal University, 200241 Shanghai, China

[b]Key Laboratory for Computational Physical Sciences (MOE), Institute of Computational Physics, Fudan University, Shanghai 200433, China

**Abstract**

The present study introduces a renormalization-based approach to investigate the relaxation dynamics within supercooled liquids. By applying a numerical scale transformation to potential energies along the temporal axis, we have established a novel framework that elucidates the underlying kinetics of supercooled liquids. Our findings indicate that the skewness of the potential energy distribution attains its maximum at a characteristic time scale ($\Delta_{CTS}$), which exhibits a scaling law with temperature $\left(\Delta_{CTS} \propto \frac{1}{(T-T^*)^{\gamma}}\right)$. This scaling relationship is characterized by an exponent ($\gamma$) that experiences a discontinuous transition at a critical cooling rate, signifying a kinetic-like phase transition. We further find that the product of $\gamma$ and the logarithm of the cooling rate is approximately constant. This constant, however, varies depending on whether the cooling rate is above or below the critical value, effectively classifying supercooled liquids into two regimes: glass-forming regime (GFR) and crystal-forming regime (CFR). Furthermore, we identify that $T^*$ corresponds to the glass transition temperature for GFR $\left(T_g\right)$ and the crystallization temperature for CFR $(T_c)$, respectively. We have successfully developed a theoretical model, which not only derives the scaling law but also provides profound insights into the physical implications of $\Delta_{CTS}$, $\gamma$ and $T^*$. This research delineates the differences between GFR and CFR, and offers a fresh perspective for exploring the nature of glasses. The findings contribute to the broader understanding of the dynamics of supercooled liquids and the mechanisms of glass formation.

[*]ZB mainly for the numerical calculations, ZDM mainly for the model analysis.
[§]Corresponding Authors: xggong@fudan.edu.cn; dysun@phy.ecnu.edu.cn

# I. Introduction

Supercooled liquids or glasses undergo a complex array of kinetic relaxation processes that span vast temporal scales [1-4]. The typical processes include β-relaxations and α-relaxations. β-relaxations, slightly more extended than atomic vibrations, pertain to relaxations as the system navigates minor energy barriers [5-9]. In contrast, α-relaxations, typically more prolonged, involve atomic rearrangements over larger energy barriers. As the temperature approaches the glass transition temperature ($T_g$), α-relaxations experience a significant increase, often characterized by a super-Arrhenius behavior [10-15]. These atypical increases in α-relaxations contribute to anomalous thermodynamic behaviors, further complicating the understanding of glasses and glass transitions.

The typical characteristics of supercooled liquids or glasses manifest not only in the anomalous temperature dependence, but also in the peculiar sensitivity to cooling rates [16-18]. An important and valuable question is whether some interesting scaling relationships are hidden in the process of glass transitions. Scaling behavior has been a staple in traditional statistical physics, serving to uncover universal principles and critical phenomena [19-22]. Consequently, examining the scaling behaviors under supercooled liquids is fundamentally essential for demystifying the nature of glasses and the glass transition. A few theories have anticipated a critical behavior in supercooled liquids [23, 24]. However, to our best knowledge, the previous theories have omitted the effect of cooling rates, potentially a pivotal aspect of the glass transition.

In this study, we propose a computational framework designed to elucidate the possible scaling behavior of the relaxation dynamics within supercooled liquids. Drawing inspiration from renormalization techniques typically applied in spatial dimensions, our methodology has been tailored for temporal analysis. Recognizing that relaxation phenomena are inherently linked to fluctuations in potential energy, the distribution of potential energies (DPE) may exhibit some unknown scaling behaviors. DPE, which encapsulates information on the local potential energy landscape (PEL) accessible to the system, provides a valuable lens through which to examine the intrinsic properties of supercooled liquids [25-42]. By incrementally expanding the temporal scale in our numerical renormalization process, we systematically filter out transient dynamics and short-time relaxation processes, thereby isolating the persistent long-time relaxation processes that are pivotal to understanding the glass transition. By examining the resulting change in DPEs with our temporal renormalization process, we have uncovered a novel characteristic time and an associated exponent, which classifying supercooled liquids into two distinct regimes: glass-forming regime (GFR) and crystal-forming regime (CFR).

## II. Computational Details

**Systems and Molecular Dynamics Simulations:** In this study, we employed the LAMMPS package [43] to investigate four distinct supercooled liquids: aluminum, copper, $Cu_{50}Zr_{50}$ alloy, and the well-recognized Kob-Andersen 80:20 (A:B) binary Lennard-Jones mixture (BLJ) [44, 45]. To characterize the interactions between atoms in aluminum, we utilized a many-body potential with parameters derived from reference [46]. The empirical potential for copper was sourced from another reference [47]. For the $Cu_{50}Zr_{50}$ alloy, we adopted an interatomic potential from reference [48]. In the case of the BLJ model, the atoms of both types interacted via a Lennard-Jones potential, with parameters sourced from reference [49], corresponding to the $Ni_{80}P_{20}$ alloy. For aluminum, copper and $Cu_{50}Zr_{50}$, the simulation consisted of 2048 atoms in a cubic box with periodic boundary conditions applied in three directions. For BLJ, the number of atoms is 1000.

There exists a critical cooling rate ($R_c$) in almost all liquids. When the cooling rate is below the critical cooling rate, the supercooled liquid undergoes crystallization at the crystallization temperature ($T_c$). $T_c$ refers to the temperature at which there is a significant jump in energy as indicated in Fig. 7 of APPENDIX A. $T_c$ possesses a substantial dynamic component and is dependent on the cooling rate. Conversely, when the cooling rate exceeds the critical cooling rate, the supercooled liquid experiences a glass transition, during which the energy continuously evolves across the whole range of temperatures. The glass transition temperature $\left(T_g\right)$ can be determined by linearly extrapolating the variation of energy or volume with temperature at high and low temperatures. $T_g$ corresponds to the intersection of these two extrapolated lines. In this paper, except for the BLJ, $T_g$ of rest systems is determined based on the variation of volume. Similar to $T_c$, $T_g$ is also cooling rate-dependent [50-56]. We found that, for aluminum and copper, $R_c$ is approximately around $10^{11}$K/s and $10^{12}$K/s, respectively. Both BLJ and $Cu_{50}Zr_{50}$ are excellent glass-former liquids with very low critical cooling rates [57, 58]. Investigating the crystallization of BLJ and $Cu_{50}Zr_{50}$ under current computational conditions is challenging. This paper focuses solely on its glass transition process.

For each system, we selected 4 to 6 cooling rates, ranging from $10^7$K/s to $10^{13}$K/s, spanning several orders of magnitude. All systems were initially equilibrated at much higher temperature to ensure a well-defined liquid state. Then, the systems were cooled to 300K using the isothermal-isobaric ensemble (NPT) at zero pressure. During this process, different cooling rates will cause the system to deviate from equilibrium and to reside in different regions of PEL. After the cooling process, at each temperature of interest, long-time canonical ensemble (NVT) simulations are performed, in which the state obtained from NPT simulations is taken as the initial state. With this protocol, the NVT simulations incorporate the effect of the cooling rate by inheriting the state information from the NPT ensemble. All data used to calculate the distribution of potential energy (DPE) are taken from the NVT simulations (Note: for BLJ, rather than NPT, the NVT simulation is used to cool the system).

For each cooling rate, we have conducted 30 to 100 independent NPT simulations starting from different initial states. Accordingly at each temperature of interest, there are also 30 to 100 independent NVT simulations. Each NVT simulation lasts for 50 ns. Therefore, for a single temperature, the total MD time ranges from 30×50ns to 100×50ns. Thus, the total simulation time for each system is no less than 15 microseconds. To calculate DPE, the 50ns MD data is further divided into 25 segments, each approximately 2ns in length. Subsequently, the DPE is calculated in each segment. Considering the 50 independent cooling processes, we have 30×25 to 100×25 MD segment at each temperature. The final DPE is the average over all 30×25 to 100×25 MD segments, which is equivalent to the average over more than 30×25 to 100×25 ensembles.

During 50ns NVT simulations, we find that the energy hardly has any observable monotonic decrease unless crystallization occurs. This indicate that the supercooled liquid does not significantly deviates from the initial state, and stays in a metastable state or a quasi-equilibrium state. A few examples are presented Fig. 8 in APPENDIX A. As can be seen from this figure, there is indeed no discernible change in energy over 50ns. Detailed calculations indicate that, the energy drift over 50ns is even no more than $10^{-6}$ eV. Additionally, if the crystallization occurs, all MD data after the 5ns preceding crystallization will not be used to calculate DPEs.

**The Numerical Renormalization Approach**:

Our numerical renormalization approach carries out a scale transformation within the temporal dimension. This transformation is executed by averaging the physical properties across a time interval (Δ), followed by a thorough analysis of the change in the time-averaged data. Δ serves as an analogy for the scale transformation in the block spin of the Ising model, dictating the dimensions of block cells. A decreased Δ signifies a smaller size of block cells, with Δ=0 signifying the absence of renormalization. As Δ increases, the time interval utilized in renormalization expands. Any scaling characteristics could manifest in the variations of Δ.

Mathematically, the temporal renormalization process can be expressed as:

$$x_k^{(\Delta)} = \frac{1}{\Delta} \sum_{i=k\Delta}^{(k+1)\Delta} x_i^{(0)}, \tag{1}$$

where $x_i^{(0)}$ denotes the time series of any physical quantity, and $x_k^{(\Delta)}$ represents the renormalized version of $x_i^{(0)}$, i.e., the average of $x_i^{(0)}$ over Δ consecutive time steps. In above equation, we stipulate that $x_k^{(\Delta)} = x_i^{(0)}$ for Δ= 0. If $f\left(\left\{x_i^{(0)}\right\}\right)$ represents an arbitrary function of $\left\{x_i^{(0)}\right\}$, $f^{\Delta}\left(\left\{x_i^{(\Delta)}\right\}\right)$, the counterpart function of $f\left(\left\{x_i^{(0)}\right\}\right)$, is the renormalized version of $f\left(\left\{x_i^{(0)}\right\}\right)$. By considering $x_i^{(0)}$ as the spin of the $i$-th lattice in the Ising model, the aforementioned equation can be interpreted as the scale transformation of block spin. The primary objective of this transformation is to transform $f\left(\left\{x_i^{(0)}\right\}\right)$ into $f^{\Delta}\left(\left\{x_i^{(\Delta)}\right\}\right)$, a function that reflects the average of $x$ over shorter time spans.

In this paper, $x$ corresponds to the potential energy and $f\left(\left\{x_i^{(0)}\right\}\right)$ corresponds to the DPE. The process of short-time averaging, or temporal renormalization, involves calculating the average of potential energy over $\Delta$ consecutive time intervals,

$$E_{\Delta}(t) = \frac{1}{\Delta}\int_{0}^{\Delta} E(t+\tau)d\tau\,, \tag{2}$$

where $E(t)$ is the instantaneous potential energy, and $E_{\Delta}(t)$ is its average after the short-time average. This averaging modifies DPE. It is evident that the temporal renormalization acts as a filter to "purify" the contributions from various short-time vibrations and relaxations. By adjusting $\Delta$, fluctuations or relaxation processes with time scales shorter than $\Delta$ are progressively dampened. The central methodology of this paper is to ascertain how the DPE evolves with $\Delta$.

We utilized skewness to quantitatively delineate the variation of $\rho(E)$. Skewness $(S)$ is calculated as follows:

$$S = \int \rho(E)\left(\frac{E-\bar{E}}{\sigma}\right)^{3} dE\,, \tag{3}$$

where $\bar{E}$ and σ represent the mean value and the standard deviation of the DPE, respectively. Namely, $\bar{E} = \int \rho(E)EdE$ and $\sigma^2 = \int \rho(E)(E-\bar{E})^2 dE$.

## III. Numerical Results and Discussions

**The Characteristic Time ($\Delta_{CTS}$):**

The application of renormalization-like techniques along the temporal dimension serves as a powerful tool, adept at highlighting subtle distinctions in DPE across varying cooling rates. Fig. 1 illustrates the DPE for Aluminum and Copper at two distinct cooling rates. For Δ=0, the DPE exhibits a quintessential Gaussian profile, aligning with the principles of statistical physics. For different cooling rates, the primary variation is observed in the average value, reflecting a minor shift in the Gaussian peak. However, as Δ increases, the difference in DPEs between two cooling rates becomes progressively evident. As elucidated in the computational details and previous studies [59, 60], for a given Δ, the dynamic processes with time scales shorter than Δ are effectively filtered out, predominantly leaving processes with relaxation times exceeding Δ. The difference in DPEs intensifies with Δ, indicating that the differences are predominantly governed by long-time relaxation processes.

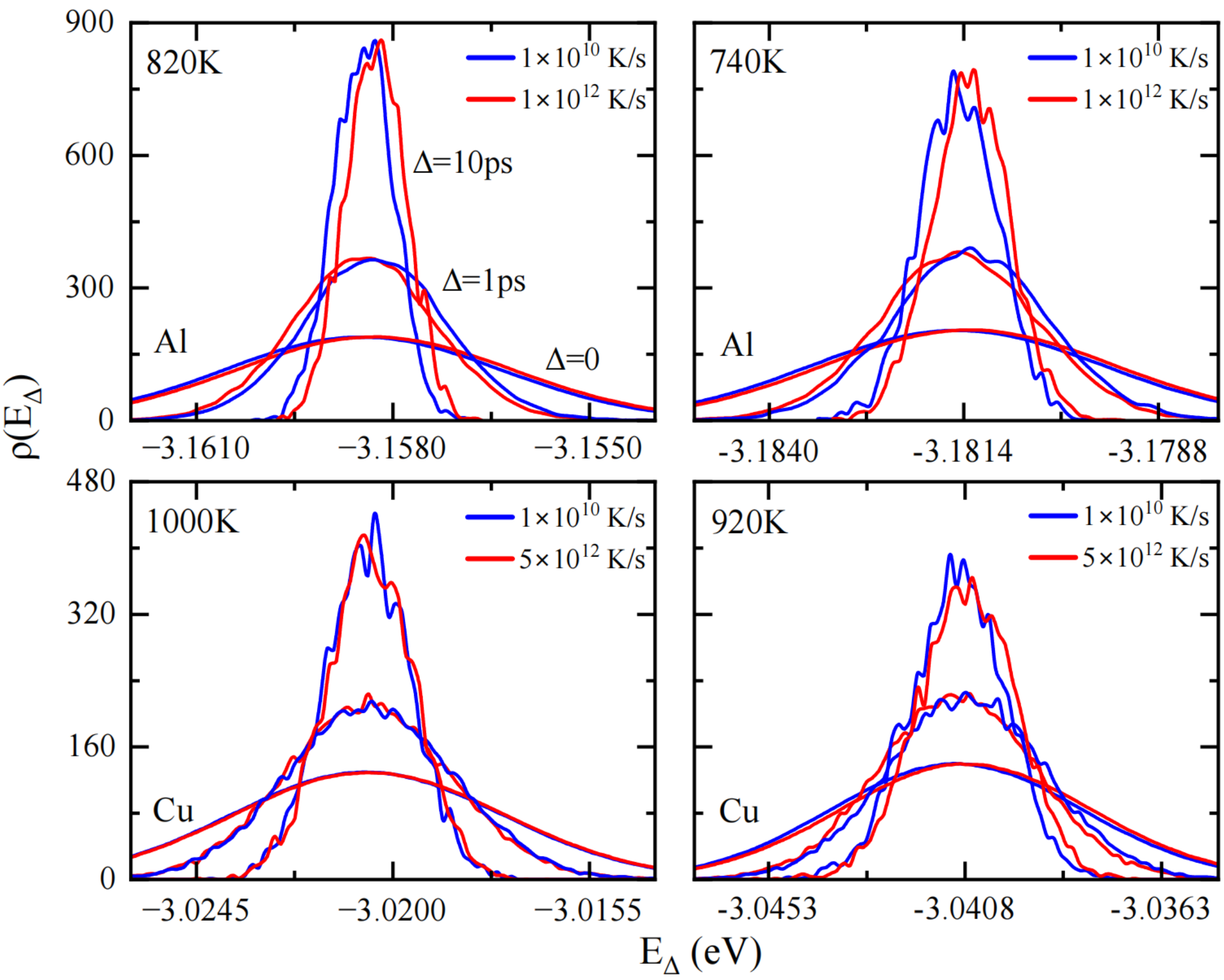

*Figure 1: Distribution of potential energies at different* Δ *and temperatures for Aluminum (upper panel) and Copper (lower panel) at two cooling rates. At* Δ= 0*, the difference in distributions is negligible. As* Δ *increases, the peak becomes increasingly higher, and the differences become more significant.*

The results shown in Fig. 1 suggest inherent differences in the DPE between the two cooling rates. To quantitatively delineate these differences, we employed skewness as a metric to characterize the variations among different cooling rates. Fig. 2 depicts the skewness variation with $\Delta$ for four different systems. It is observed that, in all examined systems, there exists a characteristic time (hereafter $\Delta_{CTS}$), where the skewness consistently attains an extremum. Moreover, the lower the temperature, the longer $\Delta_{CTS}$ is required. This outcome aligns with the anticipation, suggesting that the presence of characteristic quantities should be discernible in the variations of $\Delta$.

$\Delta_{CTS}$ emerges as a novel characteristic time, distinct in several aspects. In APPENDIX B, we have made a detailed comparison between $\Delta_{CTS}$ and other characteristic times from multiple perspectives. Here only the main conclusions are presented. Firstly, it does not conform to any existing characteristic time associated with supercooled liquids, typically lasting several times longer than the $\alpha$-relaxation time across most temperatures (see Fig. 9 of APPENDIX B for details). Secondly, $\tau_\alpha$ is almost independent of the cooling rate, while $\Delta_{CTS}$ shows a clear and strong dependence on the cooling rate. Although the cooling rates differs by three orders of magnitude, the difference in $\tau_\alpha$ is neglectable (see Fig. 10 of APPENDIX B for details). Thirdly, at the glass transition temperature or crystallization temperature, $\Delta_{CTS}$ tends towards infinity (as shown below), diverging from the behavior of the $\alpha$-relaxation time [61, 62]. In fact, the temperature dependence of $\Delta_{CTS}$ and $\tau_\alpha$ has clear differences. For instance, if we fit $\Delta_{CTS}$ and $\tau_\alpha$ separately for the same cooling rate using the scaling law (Eq. 4), the resulting $\gamma$ differ by about one-third (see Tables R1 and R2 of APPENDIX B for details). Finally, our subsequent theoretical analysis indicates that $\Delta_{CTS}$ signifies a novel characteristic time at which the fluctuation in potential energies reaches a maximum, or when the system achieves ergodicity (for details, see the section of Theoretical Model and Interpretation).

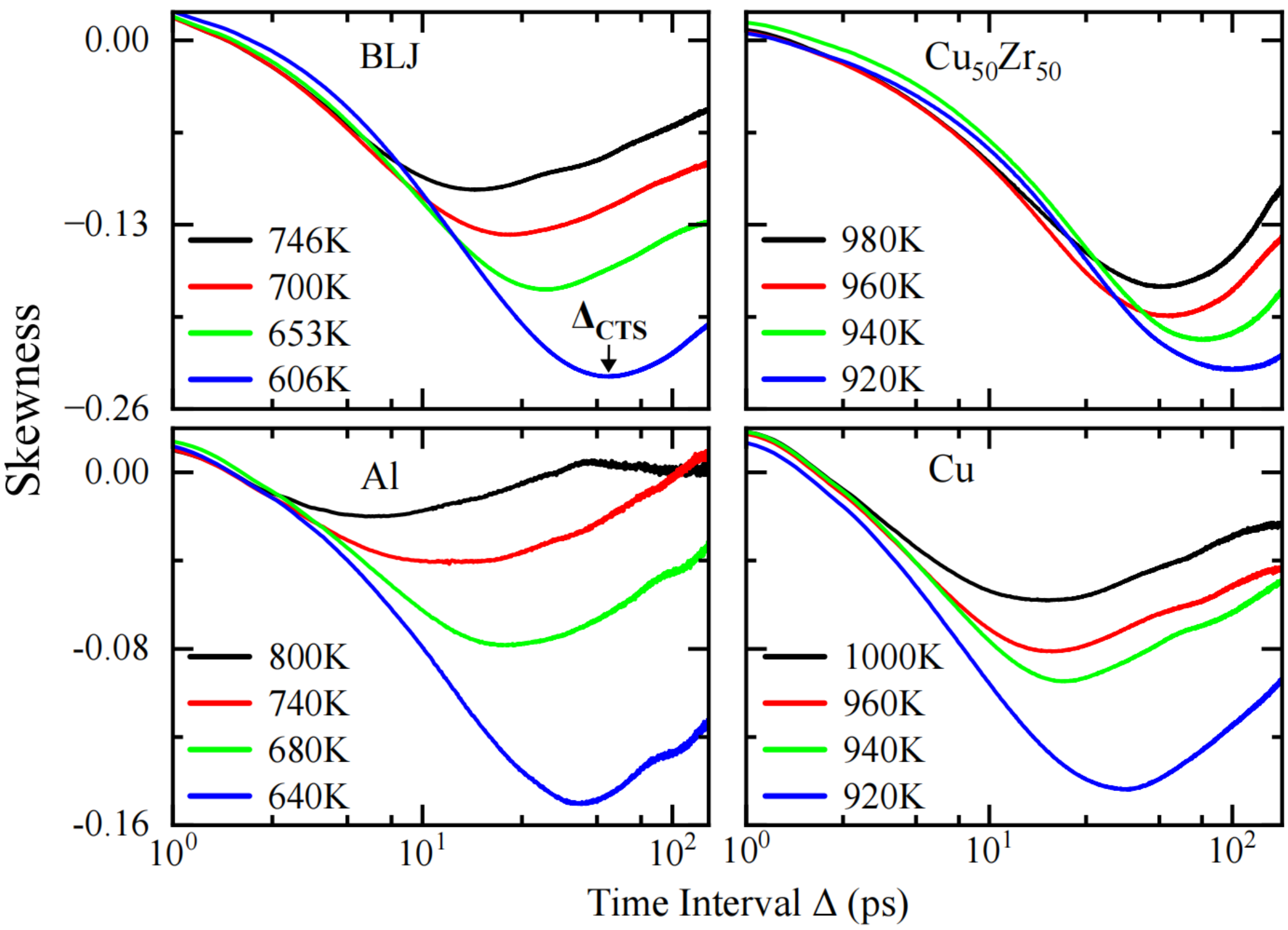

*Figure 2: The variation of skewness with the time interval* $\Delta$ *for BLJ, Aluminum,* $Cu_{50}Zr_{50}$ *and Copper for a few selected cooling rates and temperatures. For BLJ and* $Cu_{50}Zr_{50}$*, the cooling rate is* $9 \times 10^8$ *and* $1 \times 10^{11}$ *K/s, respectively. For copper, the black and red lines refer the cooling rate of* $1 \times 10^{10}$ *K/s, which the blue and green lines refer the cooling rate of* $5 \times 10^{12}$ *K/s. For aluminum, the black and red lines refer the cooling rate of* $1 \times 10^{10}$ *K/s, which the blue and green lines refer the cooling rate of* $1 \times 10^{12}$ *K/s. At all studied systems and cooling rates, a characteristic time (*$\Delta_{CTS}$*) is identified, at which skewness reaches an extremum.*

**The Scaling Law:**

For all the investigated systems, $\Delta_{CTS}$ and temperature adhere to a scaling law. Specifically, their relationship is governed by:

$$\Delta_{CTS} \propto \frac{1}{(T - T^*)^\gamma}, \tag{4}$$

where $\gamma$ and $T^*$ are constants for a given cooling rate. Fig. 3 presents the $\Delta_{CTS}$ as a function of temperatures, along with fitting results with Eq. (4). It is evident that the data aligns well with the scaling law (solid line in Fig. 3). To more clearly present the fitting results in Fig. 3 of the main text, we have plotted each system separately in Fig. 12-15 in APPENDIX B. Through this fitting, the values of $\gamma$ and $T^*$ can be determined.

In the previous, many formulas have been used for fitting relaxation time. One may wonder why Eq. (4) was chosen to describe $\Delta_{CTS}$ over other formulas. To address this question, we selected two representative formulas and conducted a detailed comparison with Eq. (4) in APPENDIX B. The two formulas are the Vogel-Fulcher-Tammann (VFT) law ($\tau = \tau_0 \exp\left[\frac{DT_0}{T - T_0}\right]$) [63-65] and the Elmatad-Chandler-Garrahan (ECG) equation ($\tau = \tau_0 \exp\left[J\left(\frac{1}{T} - \frac{1}{T_o}\right)^2\right]$) [66, 67]. As discussed in APPENDIX B, although both VFT and ECG also provide a good fit to $\Delta_{CTS}$, the scaling law provides the best fit with the smallest fitting error. More importantly, the scaling law is developed from its own theoretical model and foundation that addresses $\Delta_{CTS}$ specifically, in contrast to the other two equations which were developed specifically for $\tau_\alpha$ (see below). As discussed in the previous response, $\tau_\alpha$ and $\Delta_{CTS}$ are fundamentally different. Especially, in both VFT and ECG, each parameter has its own physical significance. $T_0$ in VFT corresponds to the ideal glass transition temperature. And $T_o$ in ECG is the onset temperature above which liquid transport has a much weaker temperature dependence. As shown in APPENDIX B, for both $T_0$ and $T_o$, the value obtained from fitting $\tau_\alpha$ and $\Delta_{CTS}$ is quite different. Therefore, $T_0$ and $T_o$ obtained by fitting to $\Delta_{CTS}$ are difficult to interpret as the ideal glass transition temperature and onset temperature, respectively. In contrast, $T^*$ in the scaling law is clearly consistent with the glass transition temperature or the crystallization temperature (see below). Thus, from a physical standpoint, the scaling law are more suitable for describing $\Delta_{CTS}$ than VFT and ECG.

Fig. 4 illustrates the variation of the characteristic exponent $\gamma$ with the cooling rate for all studied systems. It is observed that $\gamma$ of aluminum and copper decreases with increasing cooling rates, but exhibits a discontinuity at the critical cooling rate (indicated by the vertical dashed line). The critical cooling rate dictates whether the system will undergo crystallization or glass transition. The discontinuity in $\gamma$ at the critical cooling rate suggests that the supercooled liquid can be categorized into two groups: CFR and GFR. For both aluminum and copper, $\gamma$ for the CFR is smaller than that for the GFR. The discontinuous jump of $\gamma$ at the critical cooling rate also implies an inherent difference between GFR and CFR.

As mentioned in the Computational Detail, for BLJ and $Cu_{50}Zr_{50}$, studying the CFR under current computational conditions is challenging due to the much low critical cooling rate [57, 58]. Thus, the supercooled liquid of BLJ and $Cu_{50}Zr_{50}$ always falls into the category of GFR for the cooling rates examined. Consequently, no jump or discontinuity is observed in $\gamma$, and $\gamma$ simply decreases with the increase of cooling rates. This perspective implies that the jump in $\gamma$ only takes place around the critical cooling rate. Thus, $\gamma$ can be regarded as a characteristic exponent that distinguishes GFR and CFR.

It is important to note that conventional thermodynamic and kinetic measures are inadequate to distinguish CFR and GFR. In APPENDIX A, we have compared the thermodynamic (energies) and structure (radial distribution function) properties between CFR and GFR. Apart from a very minor shift in the average energy among different cooling rates, the structural information remains nearly identical at the same temperature. Moreover, in APPENDIX A and B, the traditional dynamics analysis was also used to investigate GFR and CFR. At least, the $\alpha$-relaxation is not enough to distinguish GFR and CFR (for details, see APPENDIX A and B).

From Fig. 4, it is evident that the value of $\gamma$ decreases for both GFR and CFR as the cooling rates increase, suggesting a possible scaling relation between $\gamma$ and cooling rates. Our analysis reveals that the product of $\gamma$ and the logarithm of the cooling rate remains nearly constant (hereafter denoted by $\eta$, $\eta = \gamma log_{10}R$ with $R$ being the cooling rate), as depicted in Fig. 5. More importantly, GFR and CFR have different values of $\eta$. This unique scaling relationship between $\gamma$ and cooling rate further demonstrates that $\gamma$ possesses dynamic and non-equilibrium characteristics, making it an appropriate characteristic exponent for describing non-equilibrium supercooled liquids.

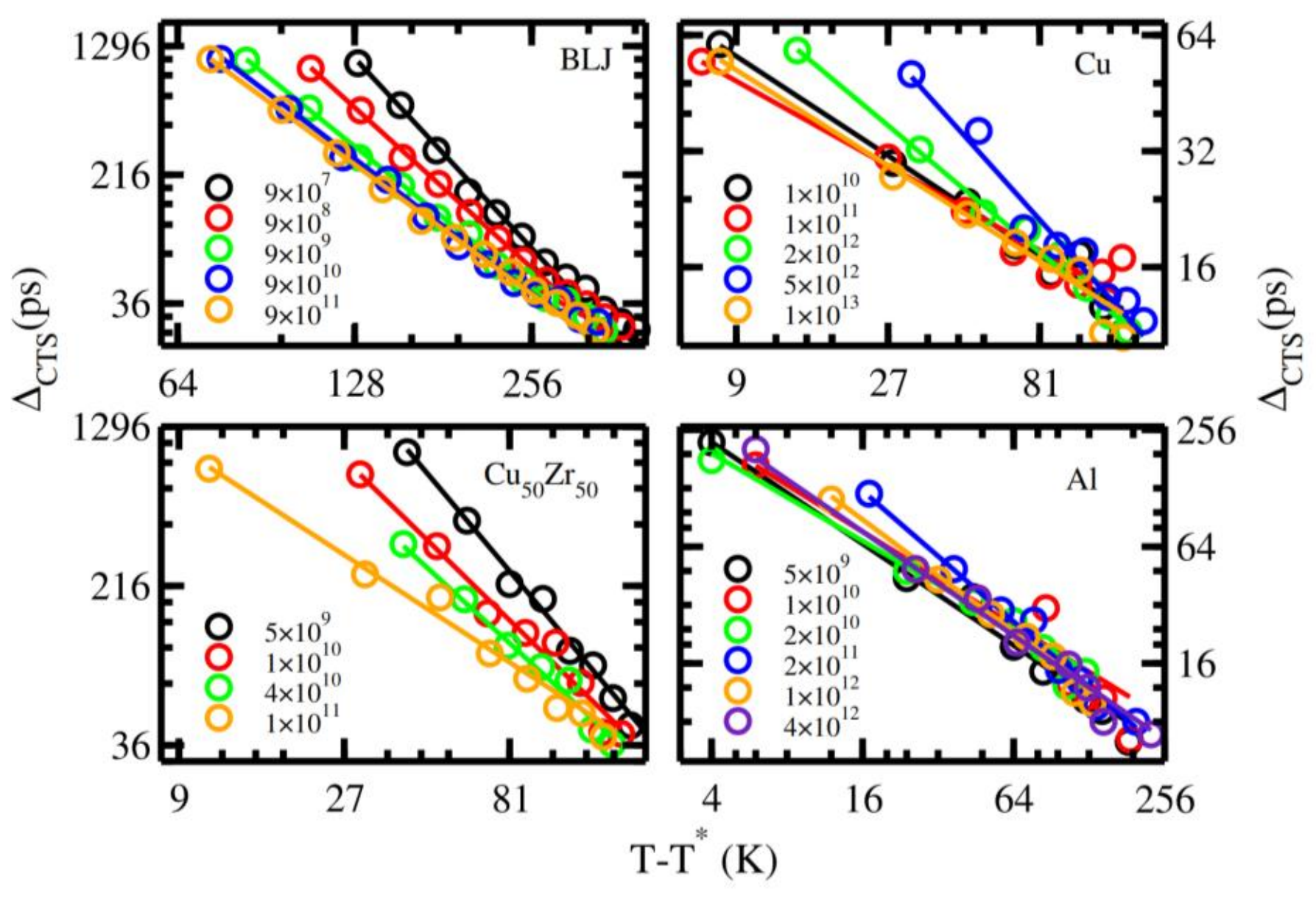

*Figure 3 : The characteristic time $\Delta_{CTS}$ via temperature for BLJ, Aluminum, $Cu_{50}Zr_{50}$ and Copper. The solid represents the best fitting with a scaling law (Eq. (4)), and $T^*$ is the fitting parameter of Eq. 4.*

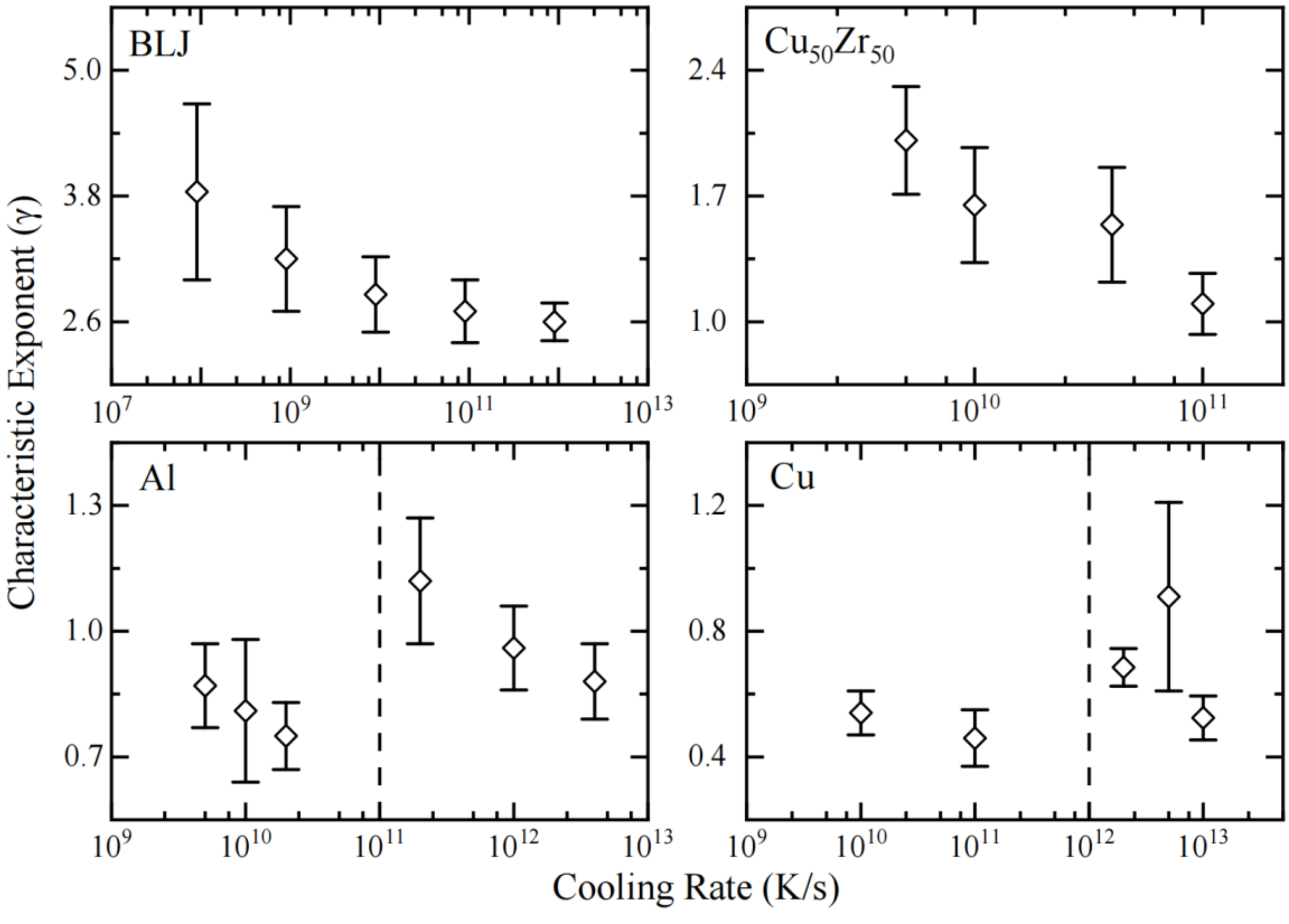

*Figure 4 : The characteristic exponent of the supercooled liquid $(\gamma)$ as a function of*

*cooling rates for BLJ, Aluminum, $Cu_{50}Zr_{50}$ and Copper. The vertical line, referring to the critical cooling rate, separates the supercooled liquid into two classes: GFR (right) and CFR (left).*

The scaling law (Eq. (4)) also predicts a critical temperature ($T^*$), at which $\Delta_{CTS}$ goes to infinity. According to our fitting results, we find that $T^*$ is nothing but the crystallization temperature ($T_c$) and the glass transition temperature ($T_g$) for CFR and GFR, respectively. It should be noted that $T_c$ is not the same as the melting point ($T_M$). $T_M$ is defined as the temperature at which the free energy of the crystal and the liquid become equal. While $T_c$ refers to the crystallization temperature for a given cooling rate, which has a significant dynamic component and depends on the cooling rate. Fig. 6 depicts $T^*$ via $T_c$ or $T_g$ obtained from MD simulations. From this figure, it is evident that, for all studied systems, $T^*$ is in good agreement with $T_c$ or $T_g$ obtained from MD simulations. This consistency should not be a coincidence, which should be a manifestation of some profound physical laws. In the next section, we will interpret $\gamma$ and $T^*$ from the perspective of statistical physics. Our analysis reveals that $\gamma$ precisely reflects the differences in PELs of the supercooled liquid, while $T^*$ carries the physical significance of the dynamic trapping temperature (see next Section).

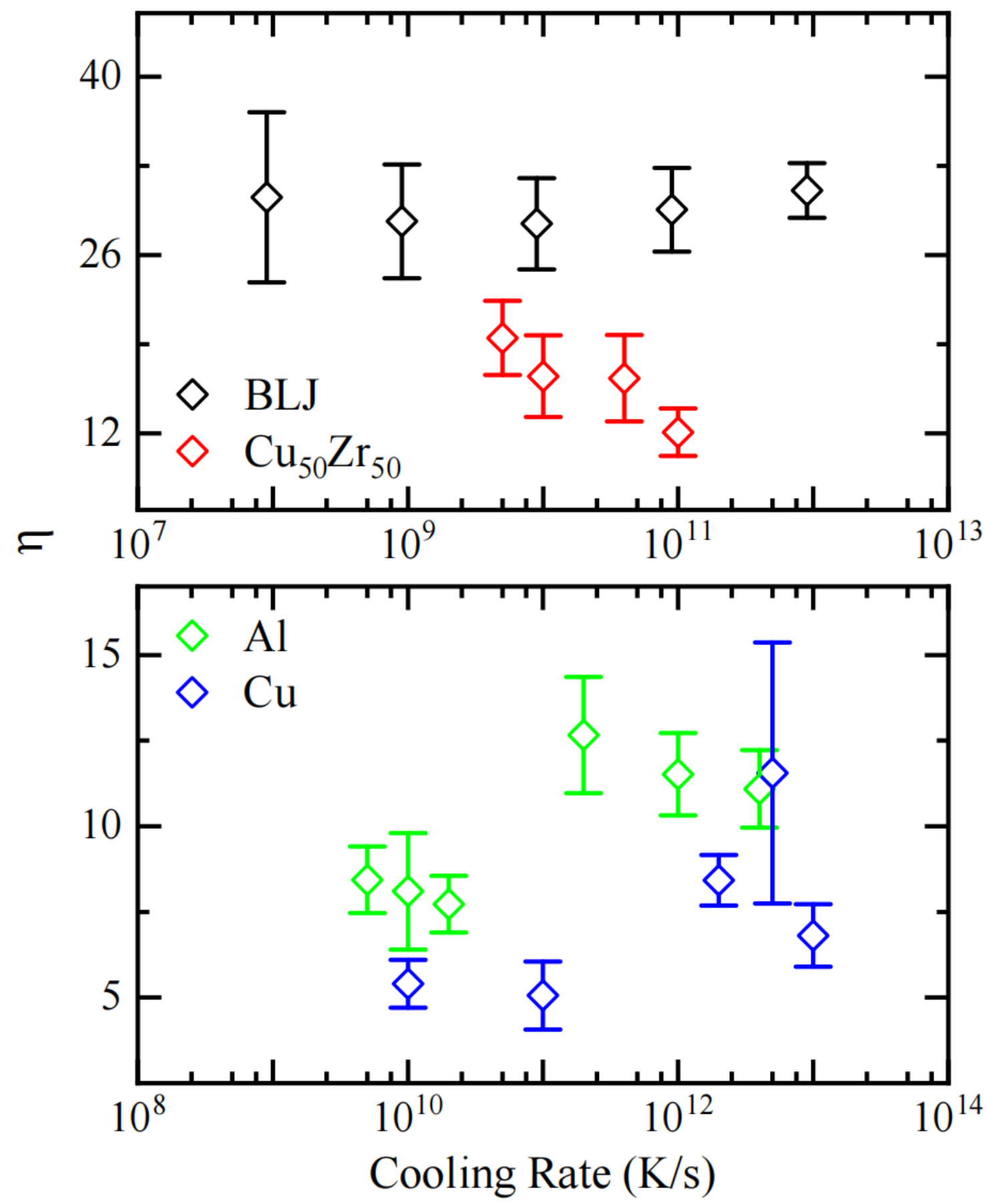

*Figure 5 : The product (η) of the characteristic exponent (γ) and the logarithm of the cooling rate for all studied systems. η almost remains constant but has different values for GFR and CFR.*

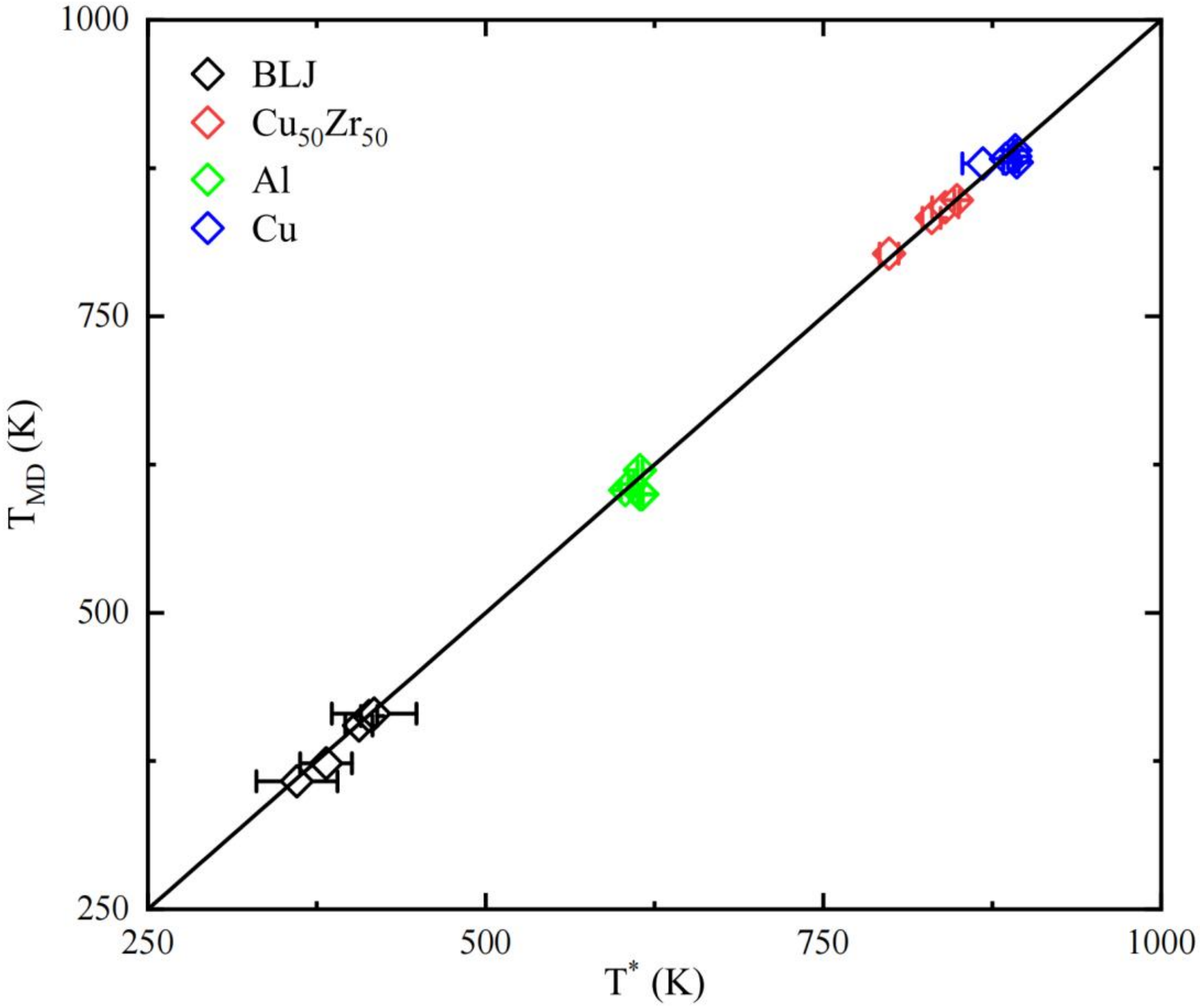

*Figure 6: The critical temperature ($T^*$) predicted by the scaling law via the glass transition temperature or the crystallization temperature obtained from MD simulations.*

## IV. Theoretical Model and Interpretation

The prevailing understanding suggests that supercooled liquids reside within a PEL with numerous metastable basins (MBs) [68-71]. Transitions between these basins and between MBs are known as $\beta$- and $\alpha$-relaxation, respectively. Our aim is to formulate a model grounded in the MB perspective within PELs, and elucidating the significance of $\Delta_{CTS}$, $\gamma$ and $T^*$, as well as the scaling law.

Considering a local PEL with $N$ MBs, each with an average potential energy $E_i$ ($i = 0,1,2 \ldots N$), statistical physics anticipates that if the system dwells significantly within the $i$-th MB, an approximate equilibrium will be achieved. Consequently, the local partition function of the $i$-th MB can be written as:

$$Z_{\mathrm{MB}_i} = \int_{E \in \mathrm{MB}_i} \rho(E) e^{-\beta E} dE \,. \tag{5}$$

At the equilibrium, the energy distribution is expected to follow a Gaussian distribution. Let $G_i(E)$ be the energy distribution and $D_i$ the number of microscopic states within the $i$-th MB, we have:

$$Z_{\mathrm{MB}_i} \approx D_i e^{-\beta E_i} \int G_i(E) dE \,. \tag{6}$$

and,

$$G_i(E) = \left(\sqrt{2\pi}\sigma_i\right)^{-1} \exp\left\{-\frac{(E - E_i)^2}{2{\sigma_i}^2}\right\}, \;\; i = 0, 1,\, 2, \ldots, N, \tag{7}$$

where $\sigma_i$ is the standard deviation (or Gaussian width) for the $i$-th MB. According to statistical physics, $\frac{{\sigma_i}^2}{k_B T^2}$ is nothing but the specific heat $(C_i)$ of the system within the $i$-th MB. $E_i$ and $k_B ln(D_i)$ are the average potential energy and the entropy of the $i$-th MB, respectively. The probability of the system occupying this MB $(P_i)$ is given by:

$$P_i = \frac{1}{Z} D_i e^{-\beta E_i}, \;\; \text{with } Z = \sum D_i e^{-\beta E_i} \,. \tag{8}$$

$Z$ is the partition function or the normalization factor. With above definitions, the overall energy distribution is:

$$\rho(E) = \frac{1}{Z} \sum_{i=1}^{N} P_i G_i(E) \,. \tag{9}$$

As detailed in APPENDIX C, the skewness $S$ is derived as:

$$S = \int \rho(E) \left(\frac{E - \bar{E}}{\sigma}\right)^3 dE = \frac{\sum_i^N P_i\left\{(E_i - \bar{E})^3 + 3{\sigma_i}^2 (E_i - \bar{E})\right\}}{\{\sum_i^N P_i (E_i - \bar{E})^2 + \sum_i^N P_i {\sigma_i}^2\}^{\frac{3}{2}}} \,. \tag{10}$$

When our numerical renormalization in the temporal dimension applies to $E(t)$, i.e., by averaging the potential energies across a time interval $(\Delta)$, $\rho(E)$ will be transformed accordingly. When consecutive potential energies are uncorrelated, the problem simplifies significantly. In this case, if the initial energy distribution is Gaussian, it retains this form post temporal renormalization. The standard deviation $(\sigma)$ decreases inversely with $\Delta$ (see APPENDIX D). However, for supercooled liquids, temporal correlations between consecutive potential energies are inevitable. Empirical observations indicate that the standard deviation of DPE decreases with a power of $\Delta$. To maintain universality, we express the Gaussian width of DPEs for the average of $\Delta$ consecutive data points using the following scaling formulation:

$$\sigma_i^{(\Delta)} = \sigma_i \cdot \Delta^{-\frac{\theta}{2}}, \tag{11}$$

where $\theta$ is a positive number, indicative of the correlation level among potential energies. When $\theta = 2$, it implies no correlation among the data. As the correlations between successive potential energies increase, $\theta$ diminishes, denoting stronger correlations. These correlations stem from dynamic correlations among atomic motions. Since our focus lies with the total potential energy of systems rather than individual atoms, the correlation between potential energy is primarily influenced by collective motions. When collective motions occur on a larger scale, the correlation between potential energy strengthens, resulting in a smaller $\theta$. It is important to note that employing the same $\theta$ for different scenarios is a simplification but does not significantly impact our conclusions.

Within the framework of PELs, the supercooled liquid transitions between different MBs, typically involving numerous collective motions [72-74]. As the system takes longer to return to a specific MB, the value of $\theta$ decreases. Thus, $\theta$ also signifies the extent of configurational entropy accessible to the system. Greater configurational entropy corresponds to a larger number of basins, leading to a smaller $\theta$. By substituting Eq. (11) into Eq. (10), we derive the expression for:

$$S^{(\Delta)} = \frac{\sum_i^N \left\{P_i(E_i - \bar{E})^3 + 3P_i{\sigma_i}^2\Delta^{-\theta}(E_i - \bar{E})\right\}}{\left\{\sum_i^N P_i(E_i - \bar{E})^2 + \Delta^{-\theta}\sum_i^N P_i{\sigma_i}^2\right\}^{\frac{3}{2}}}. \tag{12}$$

where $S^{(\Delta)}$ represents the skewness of DPEs over $\Delta$ consecutive time steps. The extremum of skewness is determined by differentiating $S^{(\Delta)}$ with respect to $\Delta$ and setting it to zero $\left(\left.\frac{\partial S^{(\Delta)}}{\partial n}\right|_{\Delta=\Delta_{CTS}} = 0\right)$ (detailed mathematical derivations are available in APPENDIX E), yielding:

$$\Delta_{CTS}=\left(\frac{\sum_i^N P_i\sigma_i^2 \sum_i^N P_i\sigma_i^2(E_i-\bar{E})}{2\sum_i^N P_i\sigma_i^2(E_i-\bar{E})\sum_i^N P_i(E_i-\bar{E})^2-\sum_i^N P_i\sigma_i^2\sum_i^N P_i(E_i-\bar{E})^3}\right)^{\frac{1}{\theta}}. \quad (13)$$

Eq. (13) reveals that at a certain temperature ($T^*$), the denominator may approach zero, leading to infinite values for $\Delta_{CTS}$. This scenario is analogous to a phase transition, specifically a kinetic phase transition. Our subsequent discussion suggests that $T^*$ could be regarded as the temperature at which the dynamic phase transition occurs.

To simplify our model and express it mathematically, we condense the $N$ MBs into two idealized MBs while preserving the core essence. This simplification is reasonable for our problem. Because we are interested in critical behaviors, corresponding to the denominator of Eq. (13) approaching zero (or $\Delta_{CTS}$ approaching infinity). In this condition, the time spent for the system to escape an MB will also approach infinity, which means that only very few MBs contribute to $\Delta_{CTS}$.

We denote the energy (the number of configurations) of the two idealized MBs as $E_0(D_0)$ and $E_1(D_1)$. To differentiate between the two MBs, we set $E_1$ equal to $E_0+\delta$, where $\delta$ is a positive value. It is clear that $P_0=\frac{1}{Z}D_0e^{-\beta E_0}$, $P_1=\frac{1}{Z}D_1e^{-\beta E_1}=P_0\frac{D_1}{D_0}e^{-\beta\delta}$. Furthermore, it is crucial to note that $P_0+P_1=1$. Through derivation (for detailed information, refer to APPENDIX F), we ultimately obtain the expression:

$$\Delta_{CTS}\cong\left(\frac{\chi}{T-T^*}\right)^{\frac{1}{\theta}}, \quad (14)$$

where $T^*$ is determined by $P_1\sigma_1^2-P_0\sigma_0^2=0$, accordingly

$$T^*=\delta\left[k_B\ln\left(\frac{D_1\sigma_1^2}{D_0\sigma_0^2}\right)\right]^{-1}. \quad (15)$$

Eq. (14) indicate that if a supercooled liquid can explore multiple MBs within a certain time, a characteristic time $\Delta_{CTS}$ and a critical temperature $T^*$ will undoubtedly exist. Moreover, the relationship between $\Delta_{CTS}$ and temperature adheres to a scaling law. At $T^*$, $P_1\sigma_1^2=P_0\sigma_0^2$, implying that the potential energy fluctuation reaches its maximum. Since $\sigma_i$ reflects the amplitude of the potential energy fluctuation for the system within a single MB, the total fluctuation of potential energies $P_0\sigma_0^2+P_1\sigma_1^2$ will reach the maximum at $P_1\sigma_1^2=P_0\sigma_0^2$. When the temperature falls below $T^*$, the system quickly becomes trapped in one MB, signifying a kinetic phase transition with the dynamic transition temperature $T^*$.

The concept of ergodicity is frequently discussed in glass research fields (e.g. [75]). Higher skewness suggests that the system can explore a broader phase space or more MBs. $\Delta_{CTS}$ represents the average time required for the skewness to reach its extremum within a specific PEL region. When $\Delta_{CTS}$ becomes infinite, it implies that the system is incapable of exploring certain PELs. Consequently, at the critical temperature $T^*$, the ergodicity experiences a complete breakdown.

It is crucial to recognize that both GFR and CFR, regardless of their specific characteristics, exist in metastable states at the thermodynamic level. Thus, when we refer to a supercooled liquid occupying a particular region on PELs, it signifies a crucial fact: there is a potential barrier separating this region from the ground state. This barrier is not excessively high, implying that, given sufficient time, the supercooled liquid can escape this region. However, the barrier should also not be too low, which ensures the supercooled liquid remaining within this region for a certain duration. This is why we describe this trapping process as a dynamic phase transition.

Comparing Eqs. (4) and (14), we observe that $\gamma$ is proportional to $\frac{1}{\theta}$. Our calculations reveal that $\gamma$ for CFR is smaller than that for GFR, suggesting that $\theta$ for CFR is larger than that for GFR. As discussed earlier, $\theta$ reflects the correlations among atomic motions. Hence, we can infer that the atomic motion in GFR exhibits stronger correlations, which may explain the observation of various collective motions in GFR [72-74, 76-78]. From Fig. 4, it appears that the value of $\gamma$ for alloys ($Cu_{50}Zr_{50}$ and BLJ) is higher than that of simple metals (Aluminum and Copper). This implies that the correlation between atomic motions in alloys is generally stronger compared to simple metals. This finding may also help to explain why alloys are more prone to forming glasses than elemental materials [53, 54].

## V. Conclusion and Outlook

In this paper, we introduce a novel computational approach known as renormalization along the temporal dimension. This method permits the progressive elimination of the impact of short-term vibrations and relaxations on potential energy, thereby enabling the extraction of insights into long-term relaxation processes. By employing this technique to analyze the distribution of potential energies within supercooled liquids, we have identified a new characteristic time ($\Delta_{CTS}$). This characteristic time follows a power-law relationship with temperature, reminiscent of the scaling law $\Delta_{CTS} \propto \frac{1}{(T-T^*)^{\gamma}}$. The discovery of this power law offers a robust method to categorize supercooled liquids into two types: GFR and CFR. And the value of $\gamma$ for GFR consistently exceeds that for CFR around the critical cooling rate. $\Delta_{CTS}$ signifies the time scale at which the potential energy fluctuation is most pronounced, indicative of a critical-like behavior. $T^*$ corresponds to a critical temperature associated with a kinetic phase transition. And $\gamma$ reflects the features of the accessible potential energy landscape of supercooled liquids. The essential difference between the two types of supercooled liquids lies in: the atomic motion in GFR may have a stronger correlation than that in CFR.

The methods and conclusions presented in this paper may exert a few influences on future research pertaining to glasses. Firstly, the computational technique termed "renormalization along the temporal dimension" is adept to separate short-time vibrations or relaxations from the dynamical processes in supercooled liquids, and thereby underscoring the dynamical behavior attributable to long-time relaxations. We think that it is crucial to consider long-time relaxations or rare relaxation events in supercooled liquids. However, there remains a dearth of computational methods equipped to address long-time relaxations or rare relaxation events effectively. We hope for the current technique to gain broader application in the investigation of long-time relaxation processes. Secondly, a pivotal conclusion we have arrived at is that there is a fundamental distinction between supercooled liquids subjected to cooling rates below and above the critical cooling rate. This conclusion imparts two critical insights: 1) Not all supercooled liquids are equally informative regarding the nature of the glasses; a judicious selection of the cooling rate is imperative; 2) Examining the differences between GFR and CFR may present a novel perspective for elucidating the nature of glasses. Thirdly, the cooling rate, $R$, is crucial in the glass formation process, and the existence of a critical cooling rate is a well-known fact [79-82]. More importantly, almost all the properties of supercooled liquids are more or less affected by $R$ [83-86]. Despite its importance, $R$ is seldom found in theoretical models concerning supercooled liquids, except in few semi-empirical equations [87]. This omission is not due to a lack of significance of $R$ to these models, but rather because current knowledge is insufficient to directly link $R$ with the properties of supercooled liquids. The present findings may break this impasse by offering a potential direct correlation between $R$ and supercooled liquids, which may promote the further development of relevant theoretical models.

Finally, $\Delta_{CTS}$ can be regarded as the timescale for third-order correlations in the potential energy relaxing over time. The present results raise a fundamental question about the nature of metastable states. According to statistical physics, an equilibrium state should exhibit time-translational invariance (TTI) across all time scales. TTI should be maintained among the renormalization process on any timescale. This is because that the process of time-renormalization merely gradually erases short-time dynamics without breaking TTI of any time-scale process. However, for a metastable or non-equilibrium state, there should exist long-time dynamics that drive the system to the equilibrium. Once these processes occur, they may vanish in time, thus TTI be broken. Unfortunately, these processes could be usually very slow， and are embedded within lower-order correlation functions in a trivial way. Therefore, we speculate that, there perhaps exists a notion of *n-order metastable equilibrium*, defined as TTI of correlations up to $n$-th order. This further implies that if the supercooled liquid can exist for a sufficiently long time without crystallizing, we can eliminate the cooling rate dependence entirely in lower-order correlations but not on the next higher-order correlations. This issue in detail will be discussed in future work.

**Acknowledgements:** This work was supported by the National Key Research and Development Program of China (Grant No. 2022YFA1404603), and by the National Natural Science Foundation of China (Grant No. 12274127 and 12188101).

## APPENDIX A: Traditional Analyses

Conventional structural analysis and thermodynamic calculations are incapable of distinguishing between CFR and GFR. Fig. 7 illustrates the energy (left panel) and radial distribution function (right panel) for both CFR (solid line) and GFR (dashed line) at two distinct cooling rates. From Fig. 7, it is evident that, for these two systems, despite orders of magnitude differences in cooling rates, the radial distribution functions at the same temperatures for the same systems are virtually indistinguishable. This observation suggests that it is challenging to differentiate these supercooled liquids based solely on their structural properties. Similarly, for varying cooling rates, although there are minor differences in average energy, there are no significant variations with temperature, except in the vicinity of the transition temperatures ($T_g$ or $T_c$).

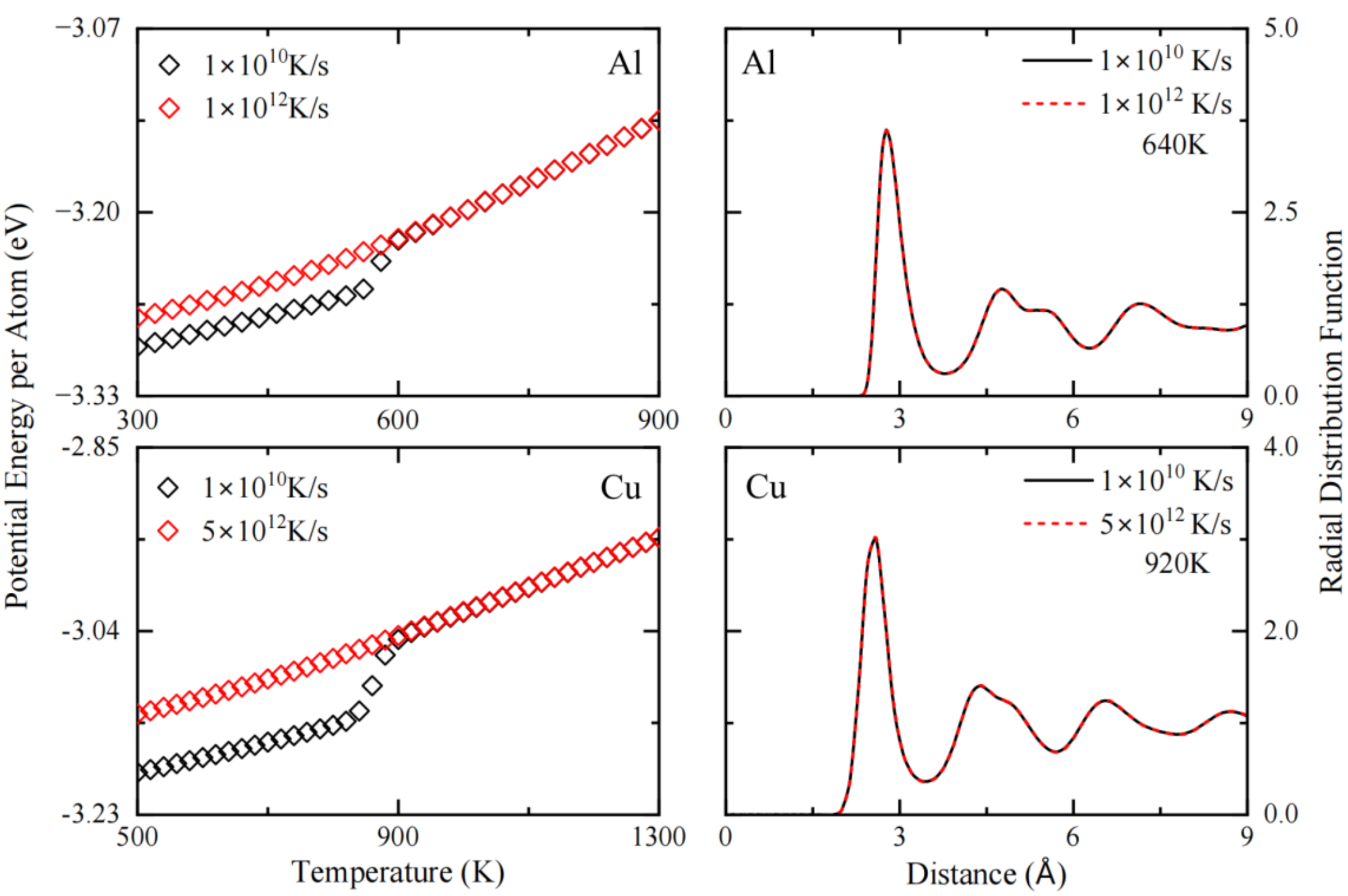

*Figure 7 : Left panel: The potential energy per atom as a function of temperature for*

*both crystal-forming liquid (solid line) and glass-forming liquid (dashed line). Right panel: The radial distribution function of crystal-forming liquid (solid line) and glass-forming liquid (dashed line) at the same temperature. The data in this figure not only includes an average over time but also over more than 200 independent cooling samples.*

In addition to the average physical quantities shown in Fig. 7, there are also no discernible evident differences in the time evolution of the physical quantities between CFR and GF. Fig. 8 presents the time evolution of energy during 50ns NVT simulations for copper and aluminum at two different cooling rates, where the higher and lower cooling rate correspond to GFR and CFR, respectively. We can see that the energy hardly has any observable monotonic decrease. This result implies that traditional dynamics analysis seems also be difficult to distinguish GFR and CFR. For examples, $\alpha$-relaxation could not be used to distinguish GFR and CFR (see below).

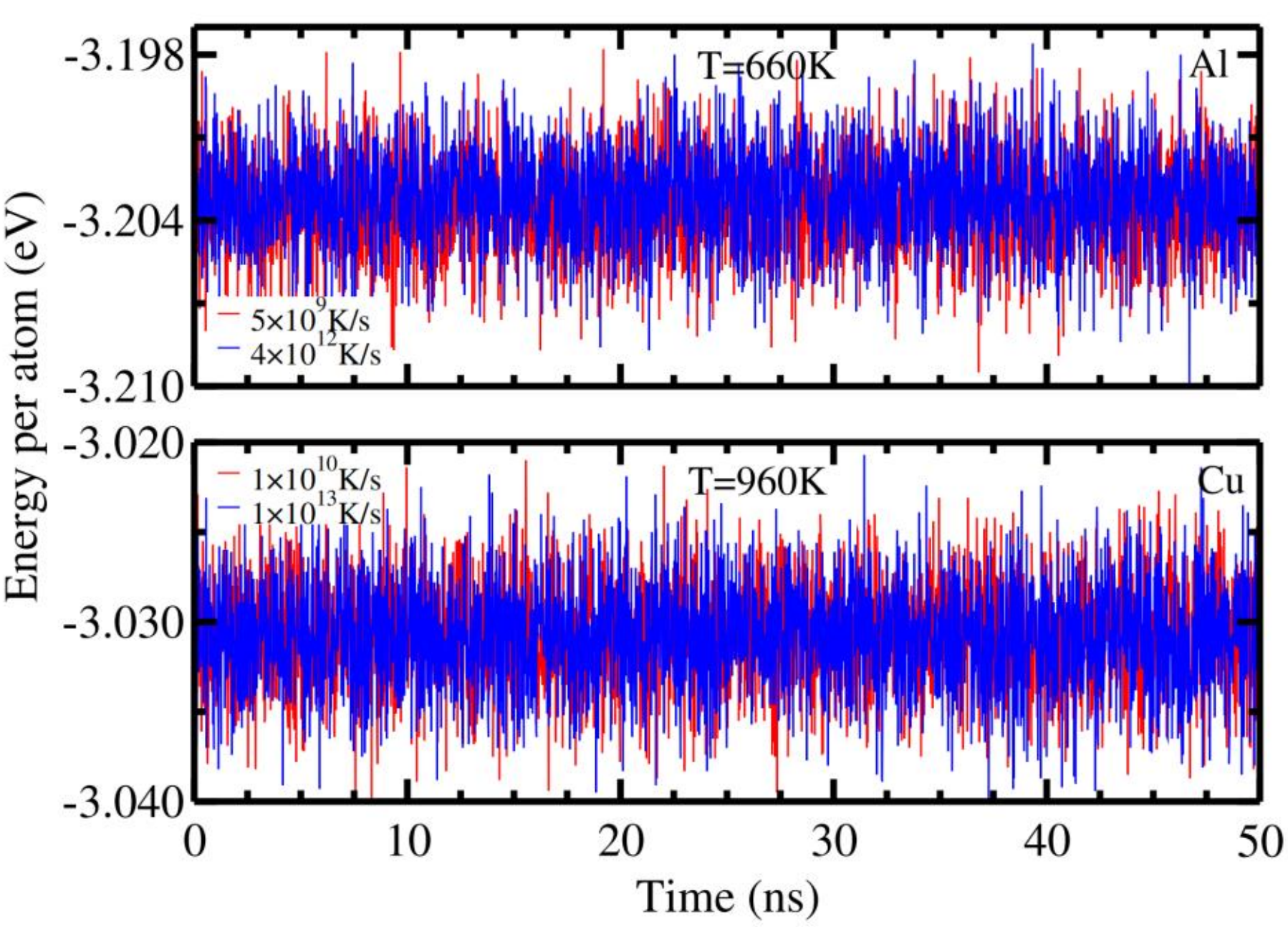

*Figure 8 : The time evolution of energy for copper (lower panel) and aluminum (upper*

*panel) at two different cooling rates. There is no discernible drift in energy over 50ns.*

## APPENDIX B: $\Delta_{CTS}$ and Scaling Law (Eq. (4))

### B.1) $\Delta_{CTS}$---a novel characteristic time

To demonstrate $\Delta_{CTS}$ is indeed a novel characteristic time, we provide a comprehensive illustration from three distinct perspectives.

First, $\Delta_{CTS}$ is more than an order of magnitude larger than both $\tau_\chi$ and $\tau_\alpha$. Here $\tau_\chi$ is the time scale at which the dynamical four-point susceptibility function $\chi_4$ reaches the maximum. The maximum of $\chi_4$ corresponds to the point in which dynamical heterogeneity, and thereby spatiotemporal correlations, are maximal. $\tau_\alpha$ is the $\alpha$-relaxation time. We have calculated $\tau_\chi$ and $\tau_\alpha$, which are depicted in Fig. 9 along with $\Delta_{CTS}$. As can be seen from this figure, for both $Cu_{50}Zr_{50}$ and BLJ, $\Delta_{CTS}$ is generally one to two order larger than $\tau_\chi$ and $\tau_\alpha$. This is sufficient to demonstrate that $\Delta_{CTS}$ is not a time associated with $\alpha$-relaxation. Of course, $\Delta_{CTS}$ is even less likely to originate from $\beta$-relaxation, which is much shorter. Although only the results for $Cu_{50}Zr_{50}$ and BLJ are shown in the figure, the conclusion is the same for the other systems studied in this work.

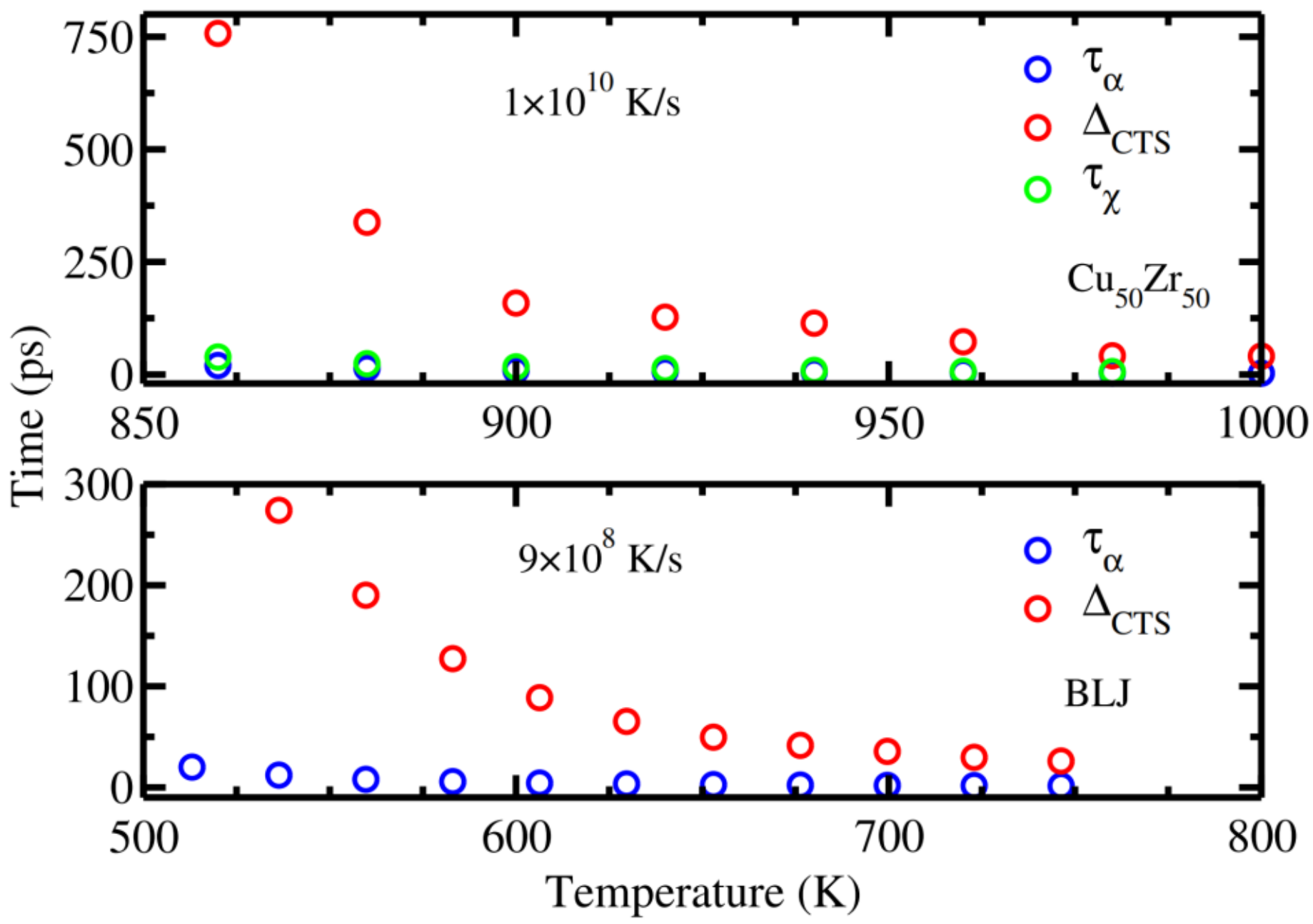

*Figure 9: α-relaxation time ($\tau_\alpha$), the time corresponding to the maximum of the dynamical 4-point susceptibility function $\chi_4$ ($\tau_\chi$), and the characteristic time ($\Delta_{CTS}$) as the function of temperatures for $Cu_{50}Zr_{50}$ (upper panel) and BLJ (lower panel).*

Second, $\tau_\alpha$ is almost independent of the cooling rate, while $\Delta_{CTS}$ shows a clear dependence on the cooling rate. Fig. 10 presents the variation of $\tau_\alpha$ with temperature for Cu at two different cooling rates. Although the cooling rates differs by three orders of magnitude, the difference in $\tau_\alpha$ is neglectable.

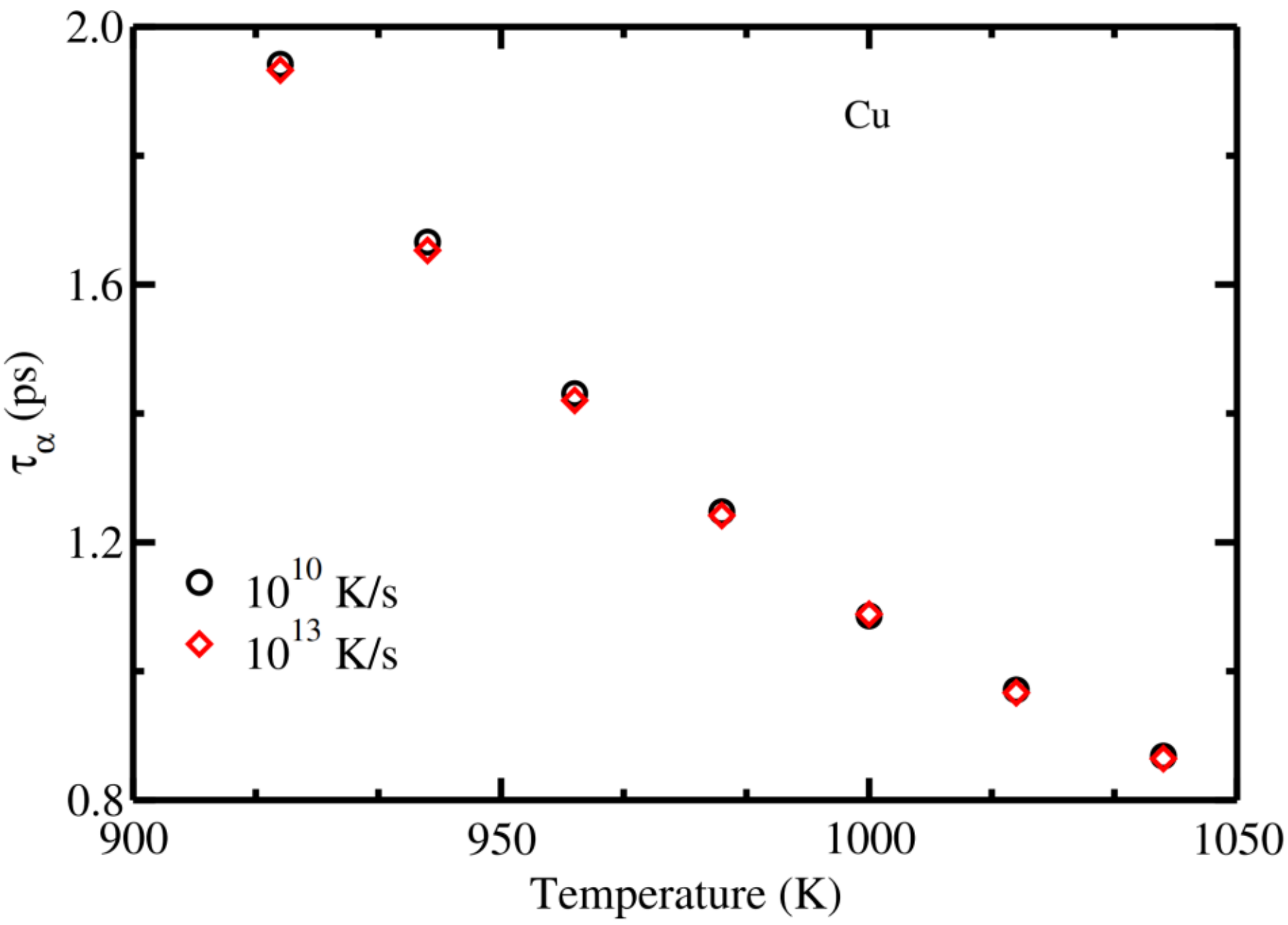

*Figure 10: α-relaxation time ($\tau_\alpha$) as the function of temperatures at two selected cooling rates for Cu. $\tau_\alpha$ is almost unaffected by the cooling rate.*

Third, the temperature dependence of $\Delta_{CTS}$ and $\tau_\alpha$ also has fundamental differences. If we fit $\Delta_{CTS}$ and $\tau_\alpha$ separately for the same cooling rate using the scaling law (Eq. 4), the resulting $\gamma$ differ by about one-third (see Tables R1 and R2). Thus, it is very difficult to attribute $\Delta_{CTS}$ and $\tau_\alpha$ to the same type of relaxation behavior.

**B.2) The scaling law more suitable for $\Delta_{CTS}$**

We have fitted $\Delta_{CTS}$ using three common used equations: the Vogel-Fulcher-Tammann (VFT) law ( $\Delta_{CTS}= \tau_0 \exp\left[\frac{DT_0}{T - T_0}\right]$), the parabolic law or the Elmatad-Chandler-Garrahan (ECG) equation ($\Delta_{CTS}= \tau_0 \exp\left[J\left(\frac{1}{T}-\frac{1}{T_O}\right)^2\right]$), and the scaling law ($\Delta_{CTS}= \tau_0 \frac{1}{(T-T^*)^\gamma}$). The fitting results are presented in Fig. 11, and the corresponding fitting parameters are listed in Table A1. Overall, all three equations provide a good fit to $\Delta_{CTS}$ data. Comparing the fitting details in Table A1, it is evident that the scaling law provides the best fitting with the smallest fitting error.

Besides of the better fitting, two more important factors incline us to choose the scaling law to describe $\Delta_{CTS}$. First, the scaling law is developed specifically for $\Delta_{CTS}$ based on a theoretical model or a theoretical foundation. In contrast, the other two equations were developed for $\tau_{\alpha}$. However, as discussed above, $\tau_{\alpha}$ and $\Delta_{CTS}$ are fundamentally different. Second, in both VFT and ECG, each parameter has its own physical significance. $T_0$ in VFT corresponds to the ideal glass transition temperature. And $T_o$ in ECG is the onset temperature above which liquid transport has a much weaker temperature dependence. By fitting to $\tau_{\alpha}$, $T_0$ and $T_o$ can be obtained, which are listed in Table A2. Comparing the data in Table A1 and A2, it is evident that, except for one case (VFT for BLJ), for both $T_0$ and $T_o$, the fitting value obtained from $\tau_{\alpha}$ and from $\Delta_{CTS}$ is quite different. Therefore, $T_0$ and $T_o$ obtained by fitting to $\Delta_{CTS}$ are difficult to interpret as the ideal glass transition temperature and onset temperature, respectively. In contrast, $T^*$ in the scaling law is clearly consistent with the glass transition temperature or the crystallization temperature. Thus, from a physical standpoint, the scaling law is more suitable for describing $\Delta_{CTS}$ than both VFT and ECG.

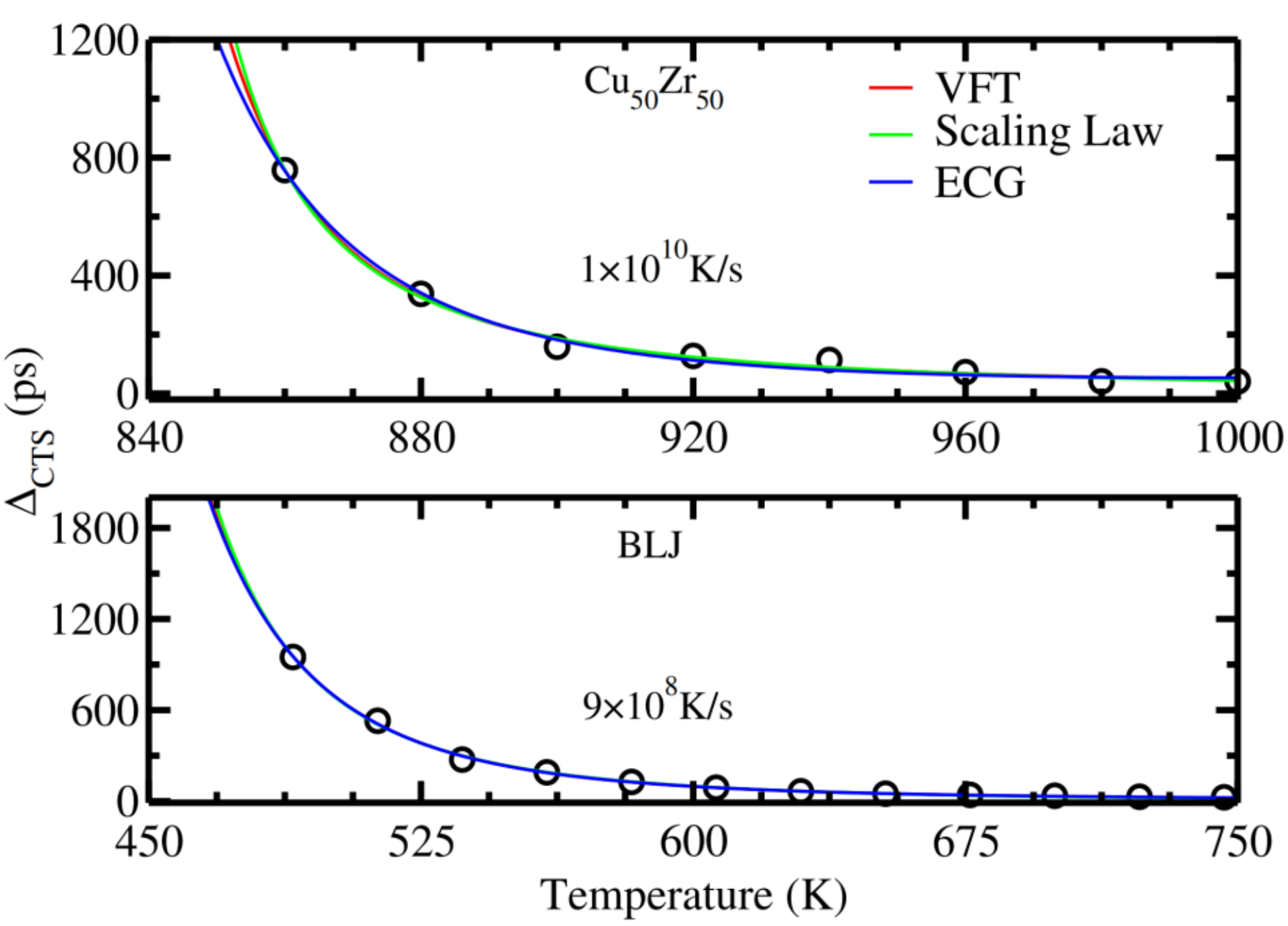

*Figure 11: Fitting to the characteristic time ($\Delta_{CTS}$) using three different formulas.*

*VFT: the Vogel-Fulcher-Tammann law,* $\Delta_{CTS}= \tau_0 exp\left[\frac{DT_0}{T-T_0}\right]$*; ECG: the parabolic law or Elmatad-Chandler-Garrahan equation,* $\Delta_{CTS}= \tau_0 exp\left[J\left(\frac{1}{T}-\frac{1}{T_0}\right)^2\right]$*; Scaling law:* $\Delta_{CTS}= \tau_0 \frac{1}{(T-T^*)^\gamma}$*. The fitting parameters are listed in Table. A1.*

*Table. A1: The parameters for VFT, ECG and Scaling Law fitting to* $\Delta_{CTS}$*. VFT: the Vogel-Fulcher-Tammann law,* $\Delta_{CTS}= \tau_0 exp\left[\frac{DT_0}{T-T_0}\right]$ *; ECG: the parabolic law or Elmatad-Chandler-Garrahan equation,* $\Delta_{CTS}= \tau_0 exp\left[J\left(\frac{1}{T}-\frac{1}{T_o}\right)^2\right]$ *; Scaling law:* $\Delta_{CTS}= \tau_0 \frac{1}{(T-T^*)^\gamma}$*.*

| | VFT | | Scaling Law | | ECG | |
|---|---|---|---|---|---|---|
| | $DT_0$ (K) | $T_0$ (K) | $\gamma$ | $T^*(K)$ | $J \times 10^{-6}(K^2)$ | $T_o\ (K)$ |
| BLJ | 1991.6 $\pm$ 507.4 | 229.3 $\pm$ 37.6 | 3.129 $\pm$ 0.47 | 385.8 $\pm$ 19.5 | 2.439 $\pm$ 0.58 | 1642.7 $\pm$ 819.0 |
| $Cu_{50}Zr_{50}$ | 406.8 $\pm$ 156.9 | 770.8 $\pm$ 20.4 | 1.648 $\pm$ 0.36 | 830.1 $\pm$ 9.5 | 100.79 $\pm$ 22.12 | 999.5 $\pm$ 24.9 |

*Table.A2: The parameters for VFT, ECG and Scaling Law fitting to* $\tau_\alpha$*. VFT: the Vogel-Fulcher-Tammann law,* $\tau_\alpha = \tau_0 exp\left[\frac{DT_0}{T-T_0}\right]$*; ECG: the parabolic law or Elmatad-Chandler-Garrahan equation,* $\tau_\alpha = \tau_0 exp\left[J\left(\frac{1}{T}-\frac{1}{T_o}\right)^2\right]$*; Scaling law:* $\tau_\alpha = \tau_0 \frac{1}{(T-T^*)^\gamma}$*.*

| | VFT | | Scaling Law | | ECG | |
|---|---|---|---|---|---|---|
| | $DT_0\ (K)$ | $T_0\ (K)$ | $\gamma$ | $T^*\ (K)$ | $J \times 10^{-6}(K^2)$ | $T_o\ (K)$ |
| BLJ | 803.1 $\pm$ 24.2 | 332.7 $\pm$ 2.4 | 2.189 $\pm$ 0.034 | 420.77 $\pm$ 1.01 | 5.36 $\pm$ 0.34 | 768.2 $\pm$ 24.8 |
| $Cu_{50}Zr_{50}$ | 1782.6 $\pm$ 835.0 | 573.2 $\pm$ 75.1 | 2.685 $\pm$ 0.69 | 737.84 $\pm$ 38.7 | 21.2 $\pm$ 6.8 | 1267.7 $\pm$ 163.6 |

To more clearly present the fitting results in Fig. 3 of the main text, we have plotted each system separately in Fig. 12-15.

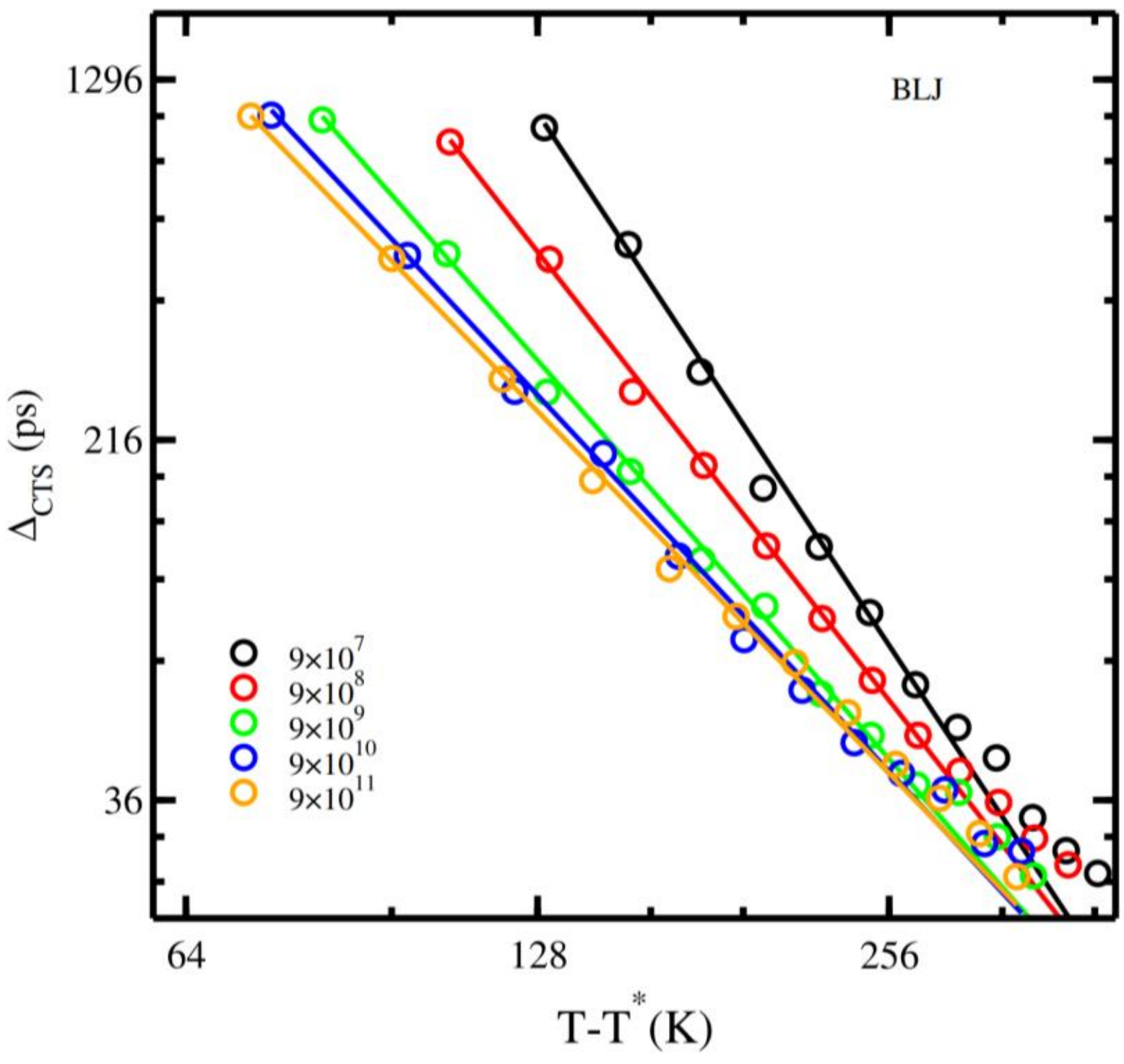

*Figure 12: The characteristic time $\Delta_{CTS}$ versus temperature for BLJ, where both the x-axis and y-axis use logarithmic scales. The solid represents the best fitting with the scaling law (Eq. (4)), and $T^*$ is the fitting value.*

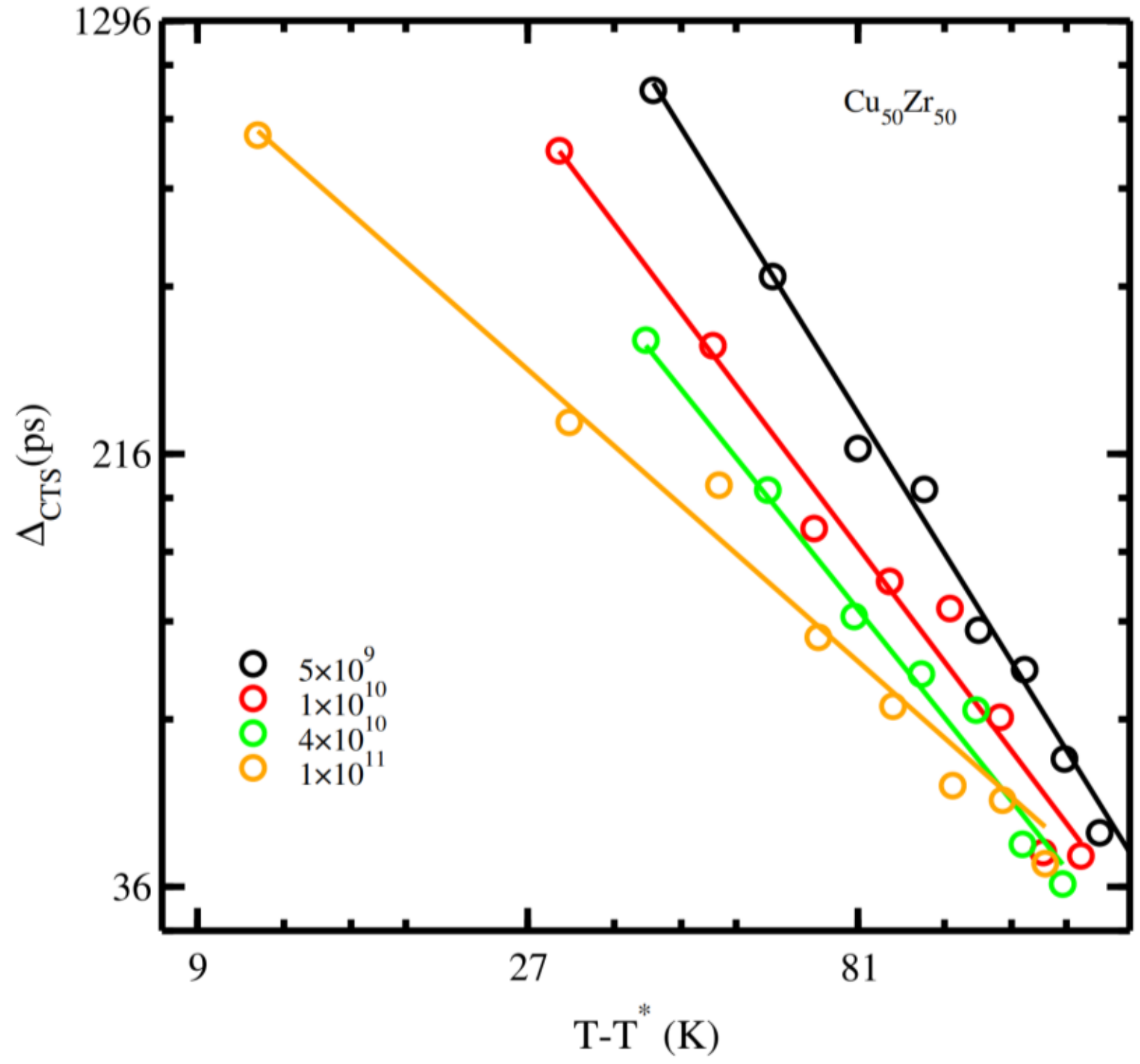

*Figure 13: The characteristic time $\Delta_{CTS}$ versus temperature for $Cu_{50}Zr_{50}$, where both the x-axis and y-axis use logarithmic scales. The solid represents the best fitting with the scaling law (Eq. (4)), and $T^*$ is the fitting value.*

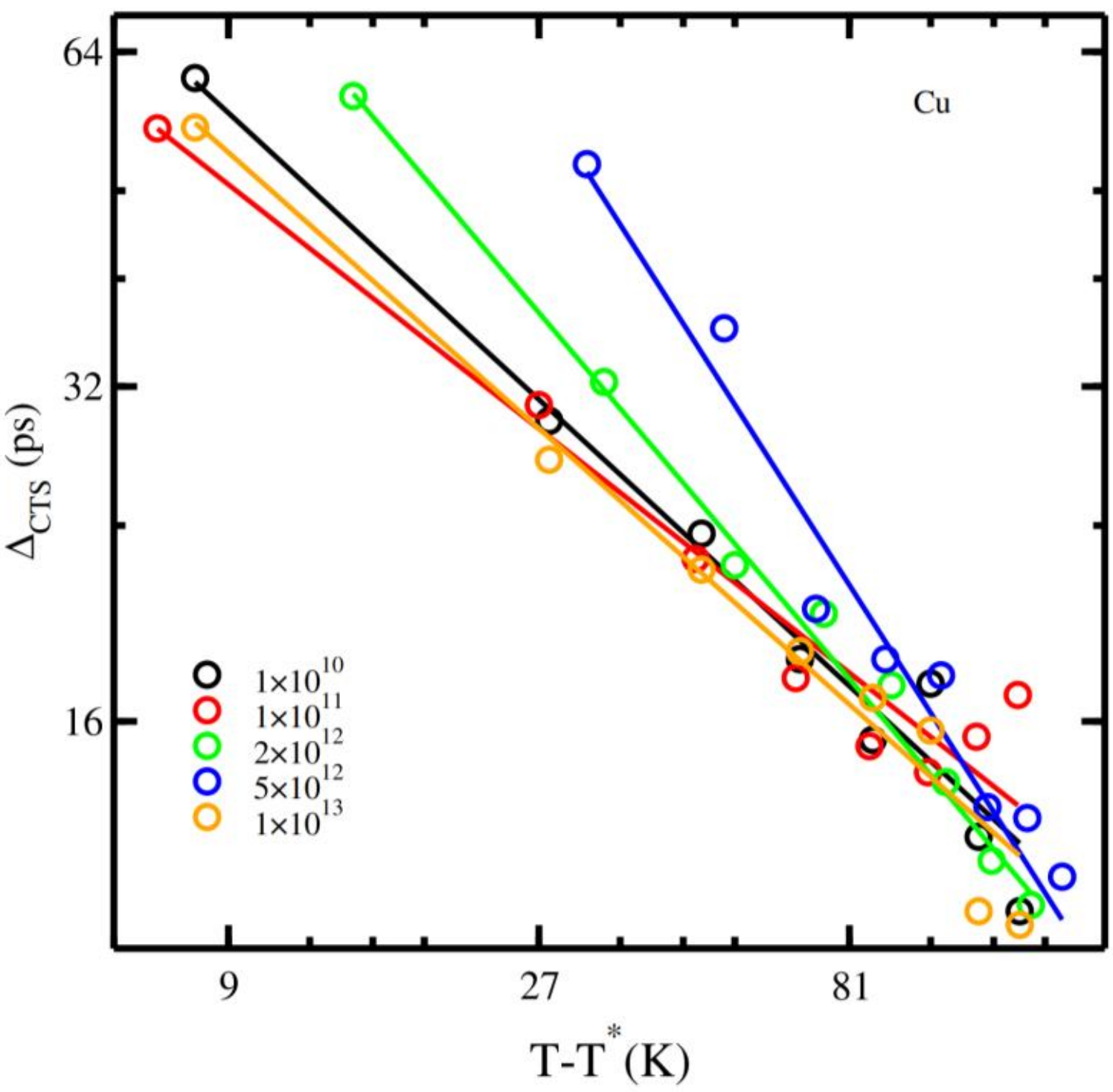

*Figure 14: The characteristic time $\Delta_{CTS}$ versus temperature for copper, where both the x-axis and y-axis use logarithmic scales. The solid represents the best fitting with the scaling law (Eq. (4)), and $T^*$ is the fitting value.*

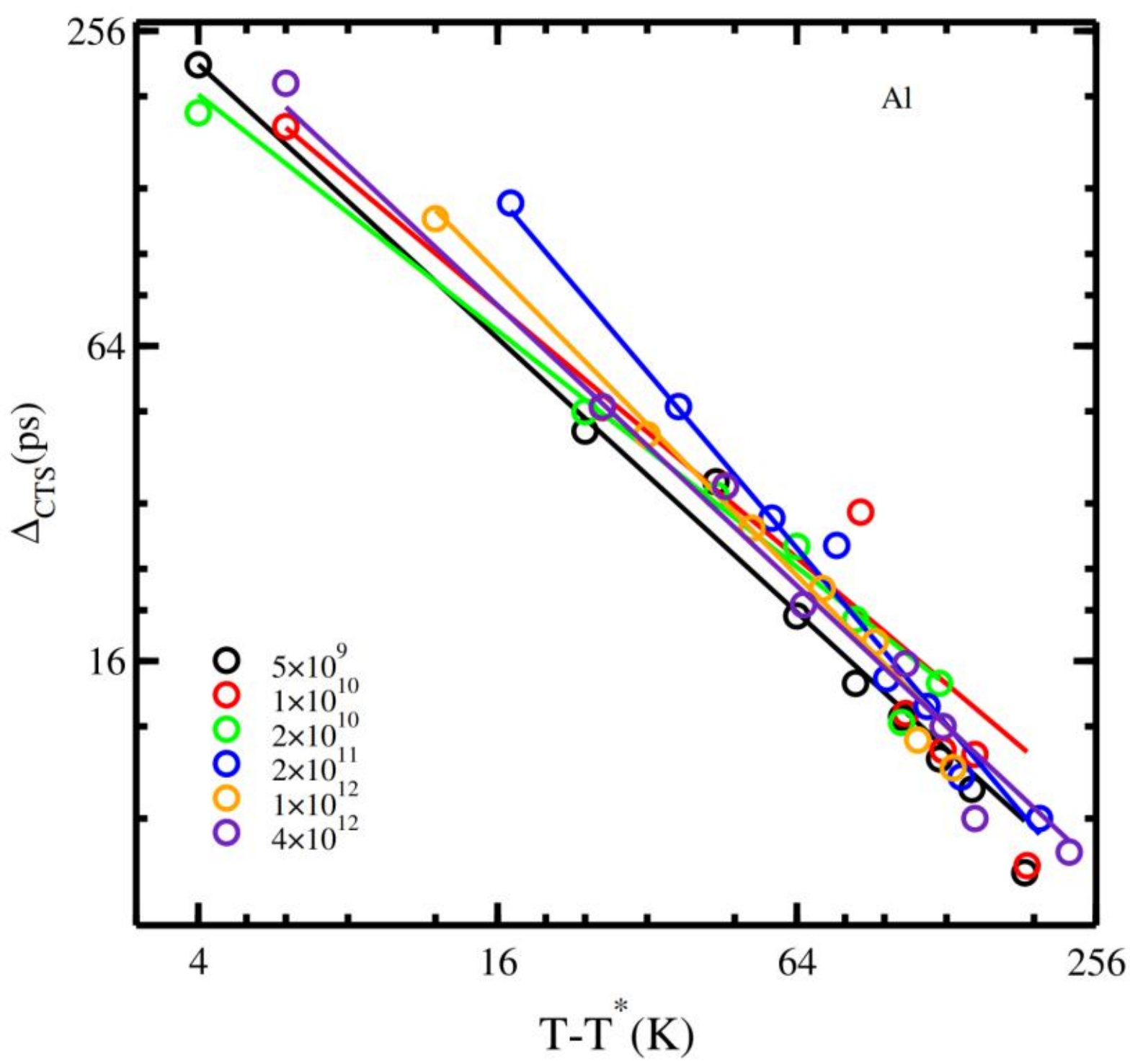

*Figure 15: The characteristic time $\Delta_{CTS}$ versus temperature for aluminum, where both the x-axis and y-axis use logarithmic scales. The solid represents the best fitting with the scaling law (Eq. (4)), and $T^*$ is the fitting value.*

## APPENDIX C: Skewness of $\rho(E)$

Considering a Gaussian distribution

$$G(x) = \left(\sqrt{2\pi}\sigma\right)^{-1} \exp\left\{-\frac{(x-\mu)^2}{2\sigma^2}\right\}$$

We have

$$\int G(x) x dx = \mu, \qquad \int G(x)(x-\mu)^2 dx = \sigma^2$$

$$\int G(x) x^2 dx = \int G(x)(x-\mu+\mu)^2 dx$$

$$= \int G(x)[(x-\mu)^2 + \mu^2 + 2\mu(x-\mu)]dx = \mu^2 + \sigma^2$$

The average potential energy of system under the distribution $\rho(E)$ in Eq. (9) in the main text reads:

$$\bar{E} = \int \rho(E) E dE$$

$\sigma$, the variance of distribution of $\rho(E)$ reads:

$$\sigma^2 = \int \rho(E)(E-\bar{E})^2 dE = \int \left[\sum_i^N P_i G_i(E)\right](E-\bar{E})^2 dE$$

$$= \sum_i^N P_i \int G_i(E)(E-\bar{E})^2 dE = \sum_i^N P_i \left[\bar{E}^2 - 2\bar{E}E_i + \int G_i(E)E^2 dE\right]$$

$$= \sum_i^N P_i\left[\bar{E}^2 - 2\bar{E}E_i + {E_i}^2 + {\sigma_i}^2\right] = \sum_i^N P_i(E_i-\bar{E})^2 + \sum_i^N P_i{\sigma_i}^2$$

$S$, the skewness of distribution of $\rho(E)$:

$$S = \int \rho(E)\left(\frac{E-\bar{E}}{\sigma}\right)^3 dE = \int \left[\sum_i^N P_i G_i(E)\right]\left(\frac{E-\bar{E}}{\sigma}\right)^3 dE$$

$$= \sigma^{-3}\sum_i^N P_i \int G_i(E)(E-\bar{E})^3 dE$$

$$= \sigma^{-3}\sum_i^N P_i \int G_i(E)[E^3 - 3E^2\bar{E} + 3E\bar{E}^2 - \bar{E}^3]dE$$

$$= \sigma^{-3}\sum_i^N P_i \left[\int G_i(E)(E-E_i+E_i)^3 dE - 3\left({E_i}^2+{\sigma_i}^2\right)\bar{E} + 3E_i\bar{E}^2 - \bar{E}^3\right]$$

in which

$$\int G_i(E)(E-E_i+E_i)^3 dE$$

$$= \int G_i(E)\left[(E-E_i)^3 + 3(E-E_i)^2E_i + 3(E-E_i){E_i}^2 + {E_i}^3\right]dE = 3{\sigma_i}^2E_i + {E_i}^3$$

Finally

$$S = \sigma^{-3}\sum_i^N P_i\left[3{\sigma_i}^2E_i + {E_i}^3 - 3\left({E_i}^2+{\sigma_i}^2\right)\bar{E} + 3E_i\bar{E}^2 - \bar{E}^3\right]$$

$$= \sigma^{-3}\sum_i^N P_i\left[{E_i}^3 - 3{E_i}^2\bar{E} + 3E_i\bar{E}^2 - \bar{E}^3 + 3{\sigma_i}^2E_i - 3{\sigma_i}^2\bar{E}\right]$$

$$= \sigma^{-3}\sum_i^N P_i[(E_i-\bar{E})^3 + 3{\sigma_i}^2(E_i-\bar{E})]$$

$$= \frac{\sum_i^N P_i\{(E_i-\bar{E})^3 + 3{\sigma_i}^2(E_i-\bar{E})\}}{\{\sum_i^N P_i(E_i-\bar{E})^2 + \sum_i^N P_i{\sigma_i}^2\}^{\frac{3}{2}}}$$

which is Eq. (10) in the main text. The process of short-time averaging or temporal renormalization involves averaging the potential energy over $\Delta$ consecutive time periods. As discussed in the main text, short-time averaging leads to:

$$\sigma_i^{(\Delta)} = \sigma_i \cdot \Delta^{-\frac{\theta}{2}}$$

Finally, we derive the Eq. (12) of the main text,

$$S^{(\Delta)} = \frac{\sum_i^N \{P_i(E_i - \bar{E})^3 + 3P_i{\sigma_i}^2\Delta^{-\theta}(E_i - \bar{E})\}}{\{\sum_i^N P_i(E_i - \bar{E})^2 + \Delta^{-\theta}\sum_i^N P_i{\sigma_i}^2\}^{\frac{3}{2}}}$$

## APPENDIX D: Gaussian Distribution without Correlation among Data

We will demonstrate that for any uncorrelated sequential data, if it adheres to a Gaussian distribution, the distribution of the mean value of any $n$ number of data points retains the Gaussian form. Moreover, the distribution width decreases with an increase in $n$. Assuming the initial distribution ($\rho_0$) has the form:

$$\rho_0(x) = a_0 e^{-b_0(x-c_0)^2}$$

The distribution $(\rho_\Delta(X))$ of the summation of any $\Delta$ data can be expressed as:

$$\rho_\Delta(X) \propto \int exp\left\{-b_0\sum_{i=1}^{\Delta}(x_i - c_0)^2\right\}\delta_D\left(X - \sum_{i=1}^{\Delta}x_i\right)dx_1 \cdots dx_\Delta$$

where $X$ is the summation of $\Delta$ data, and $\delta_D$ is the Dirac delta function. If $\delta_D$ takes the Gaussian form, namely, $\delta_D = \lim_{\kappa\to\infty} exp\left\{-\kappa\left(X - \sum_{i=1}^{\Delta}x_i\right)^2\right\}$, then the above equation becomes,

$$\rho_\Delta(X) \propto \lim_{\kappa\to\infty}\int exp\left\{-b_0\sum_{i=1}^{\Delta}(x_i - c_0)^2 - \kappa\left(X - \sum_{i=1}^{\Delta}x_i\right)^2\right\}dx_1 \cdots dx_\Delta =$$

$$\lim_{\kappa\to\infty}\int exp\left\{-b_0(x_1 - c_0)^2 - \kappa\left(x_1 - X + \sum_{i=2}^{\Delta}x_i\right)^2 - b_0\sum_{i=1}^{\Delta}(x_i - c_0)^2\right\}dx_1 \cdots dx_\Delta$$

$$= \lim_{\kappa\to\infty}\int exp\left\{-b_0(x_1-c_0)^2-\kappa\left(x_1-X+\sum_{i=2}^{\Delta}x_i\right)^2\right.$$

$$\left.-\, b_0\sum_{i=1}^{\Delta}(x_i-c_0)^2\right\}dx_1\cdots dx_\Delta$$

By integrating $x_1$ using Gaussian integral, we obtain

$$\rho_\Delta(X)\propto\int exp\left\{-b_0\sum_{i=2}^{\Delta}(x_i-c_0)^2\right\}\times$$

$$exp\left\{-\frac{b_0\kappa}{b_0+\kappa}\left(c_0-X+x_2+\sum_{i=3}^{\Delta}x_i\right)^2\right\}dx_2\cdots dx_\Delta$$

By integrating $x_2$, we arrive,

$$\rho_\Delta(X)\propto\int exp\left\{-b_I\sum_{i=4}^{\Delta}(x_i-c_I)^2\right\}\times$$

$$exp\left\{-b_0(x_3-c_I)^2-\frac{b_0\kappa}{b_0+2\kappa}\left(2c_0-X+x_2+\sum_{i=3}^{\Delta}x_i\right)^2\right\}dx_3\cdots dx_\Delta$$

This process continues until all variables being integrated, yielding:

$$\rho_\Delta(X)\propto exp\left\{-\frac{b_0\kappa}{b_0+\Delta\kappa}(\Delta c_0-X)^2\right\}=exp\left\{-\Delta b_0\left(c_0-\frac{X}{\Delta}\right)^2\right\}$$

Since $\frac{X}{\Delta}$ is the mean value $(\bar{X})$ of $\Delta$ numbers, the distribution of the mean value reads,

$$\rho_\Delta(\bar{X})\propto exp\{-\Delta b_0(c_0-\bar{X})^2\}$$

From above equation, one can see that, $\rho_\Delta(\bar{X})$ is still a gaussian distribution. And $\rho_\Delta(\bar{X})$ has the same mean value $(c_0)$ as $\rho_0(x)$, but the gaussian width reduces as $\frac{1}{\Delta}$,

$$\sigma_i^{(\Delta)}=\frac{\sigma_i}{\Delta}$$

## APPENDIX E: Reduction Eq. (13) in the Main Text

Let $f(\Delta)=\sum_i^N P_i\{(E_i-\bar{E})^3+3\Delta^{-\theta}{\sigma_i}^2(E_i-\bar{E})\}$ , $g(\Delta)=\{\sum_i^N P_i\left[(E_i-\bar{E})^2+\Delta^{-\theta}{\sigma_i}^2\right]\}^{\frac{3}{2}}$, we have

$$f'(\Delta) = -3\theta\Delta^{-\theta-1}\left[\sum_i^N P_i {\sigma_i}^2 (E_i - \bar{E})\right]$$

$$g'(\Delta) = -\frac{3}{2}\theta\Delta^{-\theta-1}\left(\sum_i^N P_i\, {\sigma_i}^2\right) g(\Delta)^{\frac{1}{3}}$$

$$\frac{dS^{(\Delta)}}{d\Delta} = \frac{g(\Delta)f'(\Delta) - g'(\Delta)f(\Delta)}{g^2(\Delta)}$$

The extreme value of skewness $S^{(\Delta)}$ is determined by taking $\left.\frac{\partial S^{(\Delta)}}{\partial \Delta}\right|_{\Delta=\Delta_{CTS}} = 0$. It requires $g(\Delta)f'(\Delta) = g'(\Delta)f(\Delta)$. After some mathematical derivation, we arrive at:

$$2\sum_i^N P_i {\sigma_i}^2 (E_i - \bar{E}) \sum_i^N P_i (E_i - \bar{E})^2 + 2{\Delta_{CTS}}^{-\theta} \sum_i^N P_i {\sigma_i}^2 \sum_i^N P_i {\sigma_i}^2 (E_i - \bar{E})$$

$$= \sum_i^N P_i\, {\sigma_i}^2 \sum_i^N P_i (E_i - \bar{E})^3 + 3{\Delta_{CTS}}^{-\theta} \sum_i^N P_i\, {\sigma_i}^2 \sum_i^N P_i {\sigma_i}^2 (E_i - \bar{E})$$

$${\Delta_{CTS}}^{-\theta} = \frac{2\sum_i^N P_i \sigma_i^2 (E_i - \bar{E}) \sum_i^N P_i (E_i - \bar{E})^2 - \sum_i^N P_i \sigma_i^2 \sum_i^N P_i (E_i - \bar{E})^3}{\sum_i^N P_i \sigma_i^2 \sum_i^N P_i \sigma_i^2 (E_i - \bar{E})}$$

which is Eq. (13) in the main text.

## APPENDIX F: Derivation of Eq. (14) in the Main Text

Let $E_1 = E_0 + \delta$, $P_0 = \frac{1}{Z} D_0 e^{-\beta E_0}$, $P_1 = \frac{1}{Z} D_1 e^{-\beta E_1} = P_0 \frac{D_1}{D_0} e^{-\beta\delta}$

The average energy reads:

$$\bar{E} = P_0 E_0 + P_1 E_1 = P_0 E_0 + (1 - P_0)(E_0 + \delta)$$

$$= E_0 + (1 - P_0)\delta$$

and

$$E_0 - \bar{E} = (P_0 - 1)\delta \text{ and } E_1 - \bar{E} = P_0\delta$$

In above derivations, we utilized the fact $P_0 + P_1 = 1$. With above mathematical relationships, Eq. (13) in the main text becomes:

$${\Delta_{CTS}}^{-\theta}$$

$$= \frac{-P_0^2 \sigma_0^2 (P_1\delta)^3 + P_1^2 \sigma_1^2 (P_0\delta)^3 + P_0 P_1 \sigma_1^2 (P_1\delta)^2 (2P_0 + P_1)\delta - P_0 P_1 \sigma_0^2 (P_0\delta)^2 (P_0 + 2P_1)\delta}{-P_0^2 \sigma_0^4 (P_1\delta) + P_1^2 \sigma_1^4 (P_0\delta) - P_0 P_1 \sigma_0^2 \sigma_1^2 (P_1\delta) + P_0 P_1 \sigma_0^2 \sigma_1^2 (P_0\delta)}$$

$$= \frac{P_1\sigma_1^2 - P_0\sigma_0^2}{(P_0\sigma_0^2 + P_1\sigma_1^2)(\sigma_1^2 - \sigma_0^2)}\delta^2$$

Rewriting the above equation as:

$$\Delta_{CTS} = \left[\frac{\sigma_1^2 - \sigma_0^2}{\delta^2} \cdot \frac{P_0\sigma_0^2 + P_1\sigma_1^2}{P_1\sigma_1^2 - P_0\sigma_0^2}\right]^{\frac{1}{\theta}} \qquad \text{(e1)}$$

It can be seen that, the critical temperature $T^*$ is determined by $P_1\sigma_1^2 - P_0\sigma_0^2 = 0$. Thus,

$$T^* = \frac{\delta}{k_B \ln\left(\frac{D_1\sigma_1^2}{D_0\sigma_0^2}\right)}$$

Taking the Taylor expansion of Eq. (e1) near $T^*$ and retaining the leading term, we obtain Eq. (14) in the main text,

$$\Delta_{CTS} \cong \left(\frac{\chi}{T - T^*}\right)^{\frac{1}{\theta}}$$

with $\chi = \frac{\sigma_1^2 - \sigma_0^2}{\delta^2} \times \frac{D_1\sigma_1^2 e^{-\frac{E_1}{k_B T^*}} + D_0\sigma_0^2 e^{-\frac{E_0}{k_B T^*}}}{\frac{E_1}{k_B T^{*2}} D_1\sigma_1^2 e^{-\frac{E_1}{k_B T^*}} - \frac{E_0}{k_B T^{*2}} D_0\sigma_0^2 e^{-\frac{E_0}{k_B T^*}}}$.